\documentclass[aps,pre,notitlepage,nofootinbib,showpacs,longbibliography]{revtex4-1}

\usepackage{amsmath, amssymb,amsfonts,textcomp}
\usepackage{color}
\usepackage{hyperref}
\usepackage{graphicx}
\usepackage[caption=false]{subfig}

\usepackage{xfrac} % "text-style" fractions

\usepackage{colortbl}

\usepackage[executivepaper,margin=0.75in]{geometry}

\usepackage{pgfplots}
\usepackage{pgfplotstable}
\pgfplotsset{width=7cm,compat=1.6}

\begin{document}

\title{Knudsen pump inspired by Crookes radiometer with a specular wall}

\author{Tobias Baier}
\email[]{baier@nmf.tu-darmstadt.de}
\affiliation{Institute for Nano- and Microfluidics, Center of Smart Interfaces, Technische Universit\"at Darmstadt, 64287 Darmstadt, Germany}

\author{Steffen Hardt}
\affiliation{Institute for Nano- and Microfluidics, Center of Smart Interfaces, Technische Universit\"at Darmstadt, 64287 Darmstadt, Germany}

\author{Vahid Shahabi}
\affiliation{High Performance Computing (HPC) Laboratory, Department of Mechanical Engineering, Ferdowsi University of Mashhad, 91775-1111, Mashhad, Iran.}

\author{Ehsan Roohi}
\affiliation{High Performance Computing (HPC) Laboratory, Department of Mechanical Engineering, Ferdowsi University of Mashhad, 91775-1111, Mashhad, Iran.}

\begin{abstract}
A rarefied gas is considered in a channel consisting of two infinite parallel plates between which an evenly spaced array of smaller plates is arranged normal to the channel direction. Each of these smaller plates is assumed to possess one ideally specularly reflective and one ideally diffusively reflective side. When the temperature of the small plates differs from the temperature of the sidewalls of the channel, these boundary conditions result in a temperature profile around the edges of each small plate which breaks the reflection symmetry along the channel direction. This in turn results in a force on each plate and a net gas flow along the channel. The situation is analysed numerically using the direct simulation Monte Carlo (DSMC) method and compared with analytical results where available. The influence of the ideally specularly reflective wall is assessed by comparing with simulations using a finite accommodation coefficient at the corresponding wall. The configuration bears some similarity with a Crookes radiometer, where a non-symmetric temperature profile at the radiometer vanes is generated by different temperatures on each side of the vane, resulting in a motion of the rotor. The described principle may find applications in pumping gas on small scales driven by temperature gradients.
\end{abstract}

% insert suggested PACS numbers in braces on next line

% insert suggested keywords - APS authors don't need to do this
%\keywords{Rarefied gas dynamics, Knudsen Pump, Direct Simulation Monte Carlo (DSMC)}

%revtex: \maketitle must follow title, authors, abstract, \pacs, and \keywords
\maketitle

%%%%%%%%%%%%%%%%%%%%%%%%%%%%%%%%%%%%%%%%%%%%%%%%%%%%%%%%
\section{Introduction\label{sec:intro}}

The objective of this paper is to demonstrate and clarify some aspects of thermally induced gas flows beyond the continuum regime when the temperature profile is shaped essentially by wall segments with high specular reflectivity. This refers to situations where the mean free path $\ell$ of the gas molecules becomes comparable to or larger than a characteristic length scale $W$ of the geometry, such as the width of a channel or the diameter of an object placed in the gas. The Knudsen number Kn=$\ell/W$ thus becomes a suitable measure of the rarefaction of the gas. Deviations from the continuum regime start to become important at Kn${\gtrsim} 0.01$, such that the Navier-Stokes-Fourier set of transport equations has to be supplemented by temperature-jump and velocity-slip boundary conditions \citep{Struchtrup_2005, Sone_2007} within the slip flow regime ($0.01 \lesssim \mathrm{Kn} \lesssim 0.1$), or replaced by an alternative description, such as the Boltzmann equation \citep{Landau_1983_X, Reif_1965, Sone_2007}, in particular within the transition flow regime ($0.1 \lesssim \mathrm{Kn} \lesssim 10$) and the collisionless or free molecular regime ($\mathrm{Kn}{\gtrsim} 10$). When a non-homogeneous temperature field is present in the gas, usually imposed by boundary conditions at walls, a gas flow can be induced, accompanied by forces on the boundaries. Classic examples are the thermophoresis of particles in a temperature gradient \citep{Tyndall_1870, Davis_2002}, the Crookes radiometer \citep{Crookes_1876, Reynolds_1879, Maxwell_1879} and thermal transpiration through a porous material with an applied temperature gradient \citep{Reynolds_1879, Knudsen_1910}. To this day these phenomena have not lost their attraction and inspire new implementations utilizing similar setups, as well as providing theoretical insight into thermally driven gas flows.

One example is the development of Knudsen pumps or compressors which have received a renewed interest with the advent of micro-mechanical systems and the associated microfabrication technologies. Optimal performance of these devices is typically reached somewhere between the slip and transition flow regimes, such that reduced length scales allows operating them at atmospheric conditions. At such small scales Knudsen pumps are particularly appealing due to their lack of moving parts, since the gas actuation is due to thermal gradients along a channel with gas flowing from a colder to a warmer region. Practically implementable designs can be accomplished based on a periodic temperature profile along the walls of a channel \citep{Sone_1996, Young_2003, An_2014}, or by inducing thermal edge flow at an array of heated plates stacked within a channel \citep{Sugimoto_2005}. An alternative implementation was recently proposed by \citet{Donkov_2011}, where the gas flow is induced between surfaces held at different temperatures. Here, one of the surfaces is considered to be structured in a ratchet pattern, with the inclined face of the ratchet reflecting gas molecules specularly, while the rest of the boundary reflects diffusely. A similar implementation relying on a purely diffusely reflecting ratchet geometry was proposed by \citet{Wurger_2011} and analyzed further in \citep{Chen_2014, Chen_2016, Wang_2016}.

Over the years the Crookes radiometer, often referred to as light mill, has served mainly as a demonstration object and as a testing ground for the theoretical understanding of rarefied gas flow in its different regimes of Kn numbers. After its discovery and the initial proposal for its operating principle by \citet{Reynolds_1879} and \citet{Maxwell_1879}, it received a flurry of attention both experimentally and theoretically in the 1920s, mostly within the German literature put forward by the likes of \citet{Westphal_1920}, \citet{Einstein_1924}, \citet{Hettner_1924}, \citet{Sexl_1926}, \citet{Epstein_1929} and \citet{Knudsen_1930}. See \citep{Martin_2010} for an overview. In this period the close relation between the Crookes radiometer and thermophoresis of particles was pointed out \citep{Einstein_1924, Sexl_1926, Epstein_1929}, and we will later review these arguments inasmuch they pertain to the situation considered here. Similarly, the close relation between thermal transpiration in Knudsen pumps and the Crookes radiometer was already pointed out by \citet{Reynolds_1879} and has recently led to alternative forms of the light mill \citep{Yang_2014, Wolfe_2016}. Conversely, \citet{Taguchi_2015} considered the situation where vanes similar to the ones of a light mill are stacked in a channel with their planes normal to the channel direction. When one side of each vane is heated, for example by illumination as in Crookes' original design, a gas flow around the edge of the vane is induced such that holding the vanes fixed leads to a net gas flow along the channel.

\begin{figure}
\begin{center}
	\includegraphics[height=5cm]{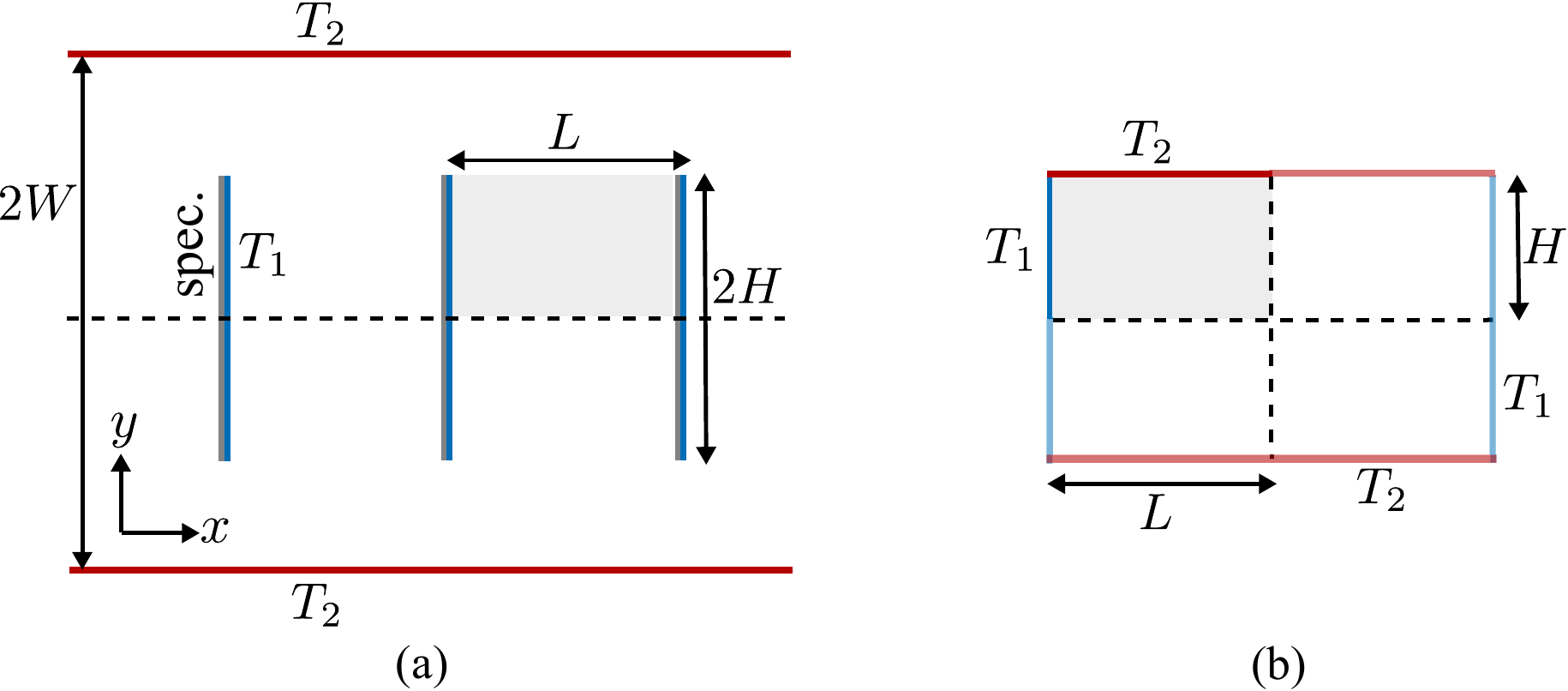}
\end{center}
\caption{(a) Sketch of the geometry. Vanes of height $2H$ are arranged with a spacing of $L$ within a channel of width $2W$. The channel walls are diffusely reflecting and held at temperature ${T}_{2}$. The left sides of the vanes are specularly reflecting while their right sides are diffusely reflecting with temperature ${T}_{1}$. (b) Simplified geometry for the force calculation at $\mathrm{Kn}={\infty}$ in section \ref{sec:collisionless}.
}\label{fig:sketch_geom}
\end{figure}

In most studies walls are considered as diffusely reflecting, and temperature gradients are imposed by walls of different temperatures in close proximity, as in the case of the Crookes radiometer with vanes having a hot and a cold side. A wall where a significant part of the impinging gas molecules does not thermalize is often considered detrimental for the operation. In this paper we investigate a situation where temperature gradients are influenced by the presence of specularly reflecting walls in an essential way. Specifically, we consider the situation sketched in figure \ref{fig:sketch_geom}a, where a series of vanes are placed inside of a channel, similar to the geometry discussed by \citet{Taguchi_2015}. Here, however, one side of each vane is considered to be ideally specularly reflecting, and a temperature difference between the diffusely reflecting side of the vane, held at temperature $T_1$, and the sidewall of the channel, at temperature $T_2$, is imposed. We will consider the vanes as fixed, thus studying the setup as a Knudsen pump. However, a similarly operated Crookes radiometer is conceivable as well. Indeed, experimentally the force on a Crookes radiometer is often quantified by operating it as a torsion balance, with the vanes suspended on a thread and measuring the deflection in a stationary state.

At first sight, studying specular surfaces may seem artificial, since it has been found experimentally that diffuse reflection represents a reasonable approximation for many surfaces, corresponding to tangential momentum accommodation coefficients not too far away from 1. However, there is clear experimental evidence that special surfaces expose accommodation coefficients much smaller than 1. For example, values between 0.1 and 0.4 have been recently reported \citep{Honig_2010, Seo_2013,Seo_2014erratum, Seo_2014, Miyoshi_2014, Lei_2015, Blanchard_2007,Cao_2009}. Therefore, the scenario considered in this paper is certainly not a standard case in terms of molecule-wall interactions, but refers to pumping principles that could become feasible with special tailor-made materials. In order to assess the influence of the ideally specularly reflective wall we will compare with simulations based on a finite accommodation coefficient at the corresponding wall. 

In the following we start by considering in section \ref{sec:collisionless} the proposed Knudsen pump with mixed specular and diffuse walls in the collisionless regime. Here, exact expressions for the forces on all boundaries can be obtained. These serve as validation of and for comparison with the numerical results obtained via Direct Simulation Monte Carlo (DSMC) in section \ref{sec:DSMC}, spanning the slip flow and transition flow regimes. In section \ref{sec:normalStress}, the results for the normal force on the vanes are qualitatively explained along the lines of an argument put forward by Einstein, extended to the range of Kn numbers considered. Section \ref{sec:MassFlux} is dedicated to the velocity field and mass transport along the channel. Additionally, the ideal condition of vanishing accommodation on one vane-face is relaxed. We close with concluding remarks in the final section \ref{sec:conclusions}.

%%%%%%%%%%%%%%%%%%%%%%%%%%%%%%%%%%%%%%%%%%%%%%%%%%%%%%%%
\section{Collisionless regime}\label{sec:collisionless}

To start with, we consider the collisionless regime, since here an analytical expression can be obtained for the force on the upper or lower wall and vanes, respectively. The force calculation proceeds along the lines of \citet{Donkov_2011}. The force density at position $\mathbf{r}_s$ on a wall is
\begin{equation}\label{eq:forceDensity}
\boldsymbol{\mathcal{F}}(\mathbf{r}_s)=-\int {m\mathbf{c} \left(\mathbf{c} \cdot \mathbf{n}\right) f ( \mathbf{r}_s, \mathbf{c} ) d^3c},
\end{equation}
where the integral is over all values of $\mathbf{c} {\in} \mathbb{R}^3$, and $\mathbf{n}$ is the unit normal on the wall, pointing into the gas phase. The phase space density for particles leaving a diffuse wall, i.e. when $\mathbf{n}\cdot\mathbf{c}>0$, is $f(\mathbf{r}_s, \mathbf{c})=n{\left(\beta/\pi\right)}^{3/2}{e}^{-\beta \mathbf{c}^{2}}$, where $\beta =m/(2k_BT)$ is determined by the temperature $T$ of the wall, and the particle number density, $n=2\sqrt{\mathit{\pi \beta }}\nu$, is determined by the particle flux density, $\nu$, (number of molecules impinging per unit time and unit area on the surface) as well as the temperature of the diffuse wall. Note that in the collisionless regime $\nu$ is constant on all walls (cf. \citet{Sone_2007}, section 2.5.1, or \citet{Hardt_2009}). At a specular wall, the phase space density obeys the symmetry relation $f ( \mathbf{r}_s, \mathbf{c} ) = f ( \mathbf{r}_s, \mathbf{c}-2\mathbf{c}\cdot\mathbf{n} )$, directly linking the distributions for impinging and reflected particles. The phase space density for particles impinging at $\mathbf{r}_s$ is obtained by tracing backwards along the particle trajectories, if necessary taking into account specular reflections, until a diffuse wall with known phase space distribution is reached. 

With these prerequisites, the average force density on the upper wall can be calculated; owing to the mirror symmetry of the model, the force density on the lower wall is obtained by mirror reflection. The calculation is simplified by noting that from the vantage point of any position within the grey area in figure \ref{fig:sketch_geom}a the upper diffuse wall looks completely homogeneous, since all particles originating from this wall have the same phase space distribution irrespective of the viewing angle. Thus the particle number, momentum or energy fluxes entering the grey area from above will be the same irrespective of the distance of the top wall from the vanes. The same is obviously also true for the fluxes leaving the grey area towards the top wall. Therefore, the analysis can be performed in the special case where $W{=}H$, such that only a single unit cell has to be considered. This is illustrated in figure \ref{fig:sketch_geom}b, where the grey area of figure \ref{fig:sketch_geom}a corresponds to the respective region in the upper left quadrant. By symmetry at the channel centre line and due to the specular side of the vane, illustrated by the horizontal and vertical dotted lines, respectively, we can extend the region of interest to one bounded by diffuse walls only, where consequently the phase space density of the outgoing particles is known. The mean momentum flux towards and away from the upper wall can thus be obtained by averaging the local force density, eq. (\ref{eq:forceDensity}), on the original section of this wall bounding the grey area. In particular, the average tangential force per area on the upper wall is
\begin{equation}\label{eq:tau_inf}
\bar{\tau}_{xy}=\frac{\sqrt{\pi}}{2}\nu m\left(\bar{c}_2 - \bar{c}_1\right)\frac{H}{L}\left[\frac{2}{\pi} \left(\arctan \frac{L}{H}-2\arctan \frac{L}{2H}\right)\right],
\end{equation}
where we have introduced the characteristic velocity $\bar{c}_i=1/\sqrt{\beta_i}=\sqrt{2k_B T_i/m}$ of diffusely reflected molecules from a wall at temperature $T_i$. The limiting values for the expression in square brackets are 0 and -1 for $L/{H}\rightarrow 0$ and ${\infty}$, respectively, and the values remain within these limits for intermediary values of $L/H$. The tangential force density vanishes for $L/{H}\rightarrow 0$ and ${\infty}$ and becomes extremal at  $L/H \approx 2.62$. We note that due to conservation of momentum a force of equal magnitude opposite to that on the sidewalls must act on the vanes, such that the corresponding average force density on a vane in $x$-direction is $\left. \bar{\mathcal{F}}_x\right|_{\mathrm{vane}} = -(L/H) \left. \bar{\tau}_{xy}\right|_{\mathrm{side}}.$

Similarly, the average normal force per area on the upper wall is
\begin{equation}\label{eq:p_inf}
 \bar{p}_{yy}=\frac{\sqrt{\pi }}{2}\nu m\left( (\bar{c}_2 +\bar{c}_1 )+( \bar{c}_2 - \bar{c}_1 )\left[\frac{2}{\pi
}\arctan \frac{L}{H}\right]\right).
\end{equation}
The limiting values for the expression in square brackets are 0 and 1 for $L/H \rightarrow 0$ and $\infty$ respectively.

\begin{figure}
	\begin{center}
		\includegraphics[scale=1]{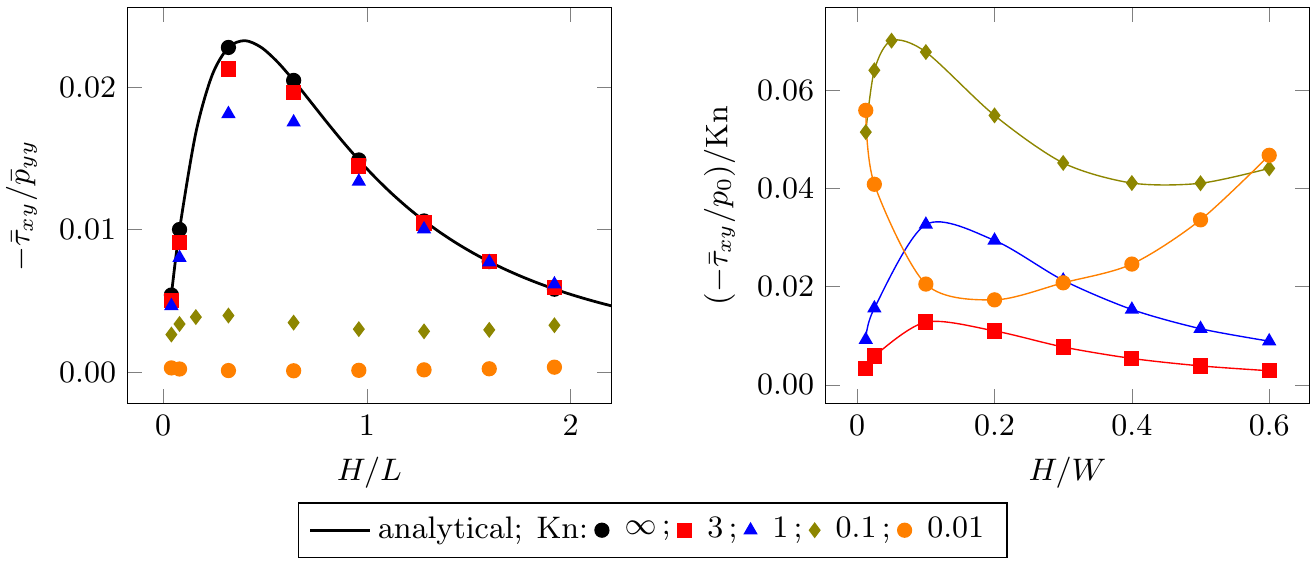}
	\end{center}
	\caption{Left: Ratio of average tangential and normal stresses on the sidewall of the channel, $\bar{\tau}_{xy}/\bar{p}_{yy}$, as function of $H/L$ for $T_2=2T_1$ and $W/L=3.2$. Solid line: analytical result in the collisionless regime, symbols: DSMC results for Kn = 0.01, 0.1, 1, 3 and $\infty$. Right: Same tangential stress data as on the left but using the pressure scale $\hat{p}_0=p_0\,\mathrm{Kn}$ for normalization, as discussed in the text (lines are merely guides to the eye).}\label{fig:force_H}
\end{figure}

The ratio between tangential and normal forces is independent of  the particle flux density $\nu$ as well as the particle mass and becomes solely a function of the geometrical parameters and the wall temperatures. As shown by the solid line on the left panel of figure \ref{fig:force_H}, the force ratio vanishes for large and small values of $H/L$ with an extremum at $H/L\approx 0.4$. In the next section we compare this result with numerical values obtained for finite Knudsen numbers. Since the net gas velocity vanishes identically in the collisionless regime \citep{Sone_2007}, obviously no mass transport takes place along the channel in this situation. Nevertheless, it is plausible that the net transfer of momentum in channel direction from the vanes to the walls results in a corresponding net mass flux as collisions between molecules become more prominent.

%%%%%%%%%%%%%%%%%%%%%%%%%%%%%%%%%%%%%%%%%%%%%%%%%%%%%%%%
\section{Finite \texorpdfstring{K\MakeLowercase{n}}{Kn} numbers}\label{sec:DSMC}

The Direct Simulation Monte Carlo Method \citep{Bird_1994} (DSMC) is a particle-based scheme for solving the Boltzmann equation for the phase space density in a rarefied gas, where particle collisions are introduced probabilistically. We use an implementation of the  method in dsmcFoamStrath \citep{Scanlon_2010, Palharini_2015, Scanlon_2015} within the framework of openFOAM \citep{Weller_1998}, version 2.3.x, relying on its capabilities of Lagrangian particle tracking. The simulation domain is one unit cell in figure \ref{fig:sketch_geom}a with $W/L=3.2$ and variable $H/L$. The vane is located in the middle of each unit cell which has cyclic boundary conditions at the left and right sides and a symmetry boundary condition at the bottom, representing the channel centreline. This domain is discretized into 100{\texttimes}300 sample cells, and the simulations are initialized such that on average 40 DSMC particles are located within each cell. During the simulation each of the cells is locally adapted to collision subcells using the transient adaptive subcell (TAS) feature of the dsmcFoamStrath solver, in a manner that the number of subcells depends on the local cell's number density \citep{Ahmad_2013}. The time step $\Delta t$ is chosen such that $\ell/{\left({ \bar{c}}_{1}\Delta t\right)}{\geq}10$ (except for Kn=0.005, where $\ell/{\left({ \bar{c}}_{1}\Delta t\right)}=5$). Each configuration was run for at least $10^6$ timesteps, where typically more timesteps were allotted to runs with lower Kn number.

For all simulations we assume that the temperature of the sidewalls is twice the temperature of the vanes, $T_2{=}2 T_1$, where the temperature scale was set to $T_1 {=} 300$~K. We define the Knudsen number as $\mathrm{Kn}{=}\ell/W$, where the mean free path $\ell{=}(\sqrt{2} {n_0} \pi d^2)^{-1}$ depends on the particle diameter and the mean particle number density $n_0$. For the DSMC simulations the variable hard sphere binary collision model with the Larsen-Borgnakke model for rotational internal energy redistribution \citep{Borgnakke_1975} is used to model collisions between molecules. We assume nitrogen with a diameter $d{=}4.17{\cdot}{10}^{-10}$~m, a viscosity temperature index of $\omega$=0.74, a reference temperature of $T_{\mathrm{ref}}{=} 273$~K, and a mass of $m{=}46.5{\cdot}{10}^{-27}$~kg. 

\begin{sloppypar}
We use the usual definitions of the particle number density, $n(\mathbf{r})={\int}f(\mathbf{r}, \mathbf{c}){d}^{3}c$, the velocity, $\mathbf{u}(\mathbf{r})={\left(n(\mathbf{r})\right)}^{-1}\int \mathbf{c}f(\mathbf{r},\mathbf{c}){d}^{3}c$ as well as the pressure tensor
 $p_{ij}(\mathbf{r})=m\int{\left(\mathbf{c}-\mathbf{u}\right)}_{i}{\left(\mathbf{c}-\mathbf{u}\right)}_{j}f(\mathbf{r},\mathbf{c}){d}^{3}c$, which is split into a diagonal and a traceless part via $p_{ij}(\mathbf{r})=p(\mathbf{r}) \delta_{ij}+\tau_{ij}(\mathbf{r})$, with the pressure $p(\mathbf{r})=(1/3)\sum_i p_{ii}(\mathbf{r})$ and the shear stress tensor $\tau_{ij}(\mathbf{r})$. The temperature obeys the ideal gas equation of state $p(\mathbf{r})=n(\mathbf{r}) k_B T(\mathbf{r})$, where ${k}_{B}$ is the Boltzmann constant. The net mass flow along the channel is the integral $\dot{m}=m \int n(\mathbf{r}) u_x(\mathbf{r}) dA$ over the cross section of the channel. For normalization we introduce ${T}_{0}{\equiv}{T}_{1}$ as the temperature scale and $\bar{c}_0 {\equiv} \bar{c}_1 {=}\sqrt{2k_B T_1/m}$ as the velocity scale, i.e. values based on the temperature at the vane. Similarly, we introduce the density scale $n_0$ as the average density corresponding to the respective Knudsen number. The corresponding pressure scale is $p_0=n_0 k_B T_0$, used for normalizing the pressure and shear stress tensors. The mass flow along the channel is normalized with $\dot{m}_0=Am n_0 c_0$, where $A$ is the full cross-sectional area of the channel. For integrated quantities such as the mass flow or area-averaged force densities, the numerical error is estimated as the standard deviation from at least three consecutive runs. For all such quantities presented in the figures within this work the error is smaller than the size of the symbols, and thus no error bars are plotted.
\end{sloppypar}

The method is validated in the collisionless regime to assess its accuracy. The absolute value of the ratio of the tangential and normal force on the sidewall of the channel is shown in the left part of figure \ref{fig:force_H} for different values of $H/L$. Since $T_2 > T_1$, the force on the sidewalls is towards the left, cf. eq. (\ref{eq:tau_inf}). Correspondingly, the force on the vanes is towards the right in this situation. As can be seen, the agreement between DSMC and analytical results is excellent. For decreasing Kn number, the ratio of tangential and normal force decreases but the functional form with respect to $H/L$ remains similar to the collisionless case up to $\mathrm{Kn} {\approx} 1$. Experiments are conveniently done in a fixed geometry by adjusting Kn via the gas pressure, and this decrease in $\bar{\tau}_{xy}/\bar{p}_{yy}$ is largely due to the increased pressure for small values of the mean free path. To directly compare forces measured at different Kn in such an experiment it is thus preferable to use a reference pressure which is independent of Kn. This is illustrated in the right panel of figure \ref{fig:force_H}, where the pressure $\hat{p}_0{=}p_0\,\mathrm{Kn}$ was used as reference scale. As can be seen, within the transition flow regime down to $\mathrm{Kn}{\lesssim}0.1$ the tangential force on the sidewall increases with increasing pressure. Its form as a function of $H/W$ also changes from its single maximum at large Kn, attained for relatively small vanes. For smaller Kn, this maximum appears to migrate towards smaller values of $H/W$, while a secondary maximum appears for vanes occupying a large part of the channel due to an increased temperature gradient between vane and sidewall, as the intermediary gap shrinks with increasing $H/W$.

\begin{figure}
	\begin{center}
		\includegraphics[scale=1]{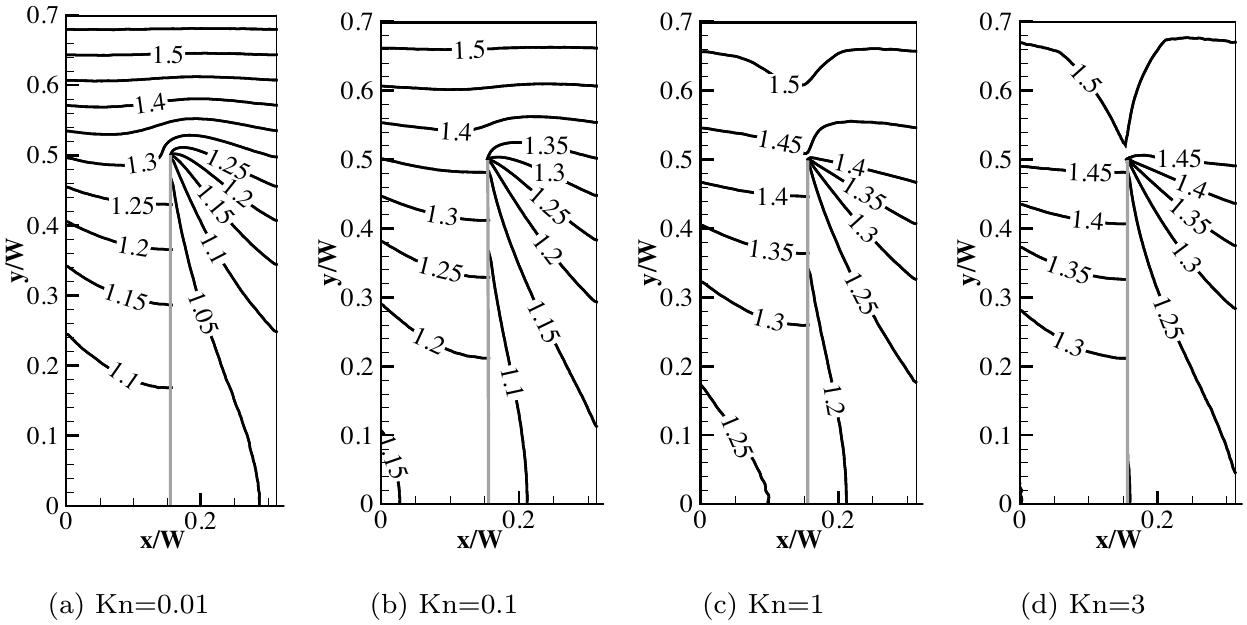}
	\end{center}	
	\caption{Isolines of temperature, $T/{{T}_{0}}$, for various Kn ($H/W=0.5$ and $H/L=1.6$).}\label{fig:iso_T}
\end{figure}

\begin{figure}
	\begin{center}
		\includegraphics[scale=1]{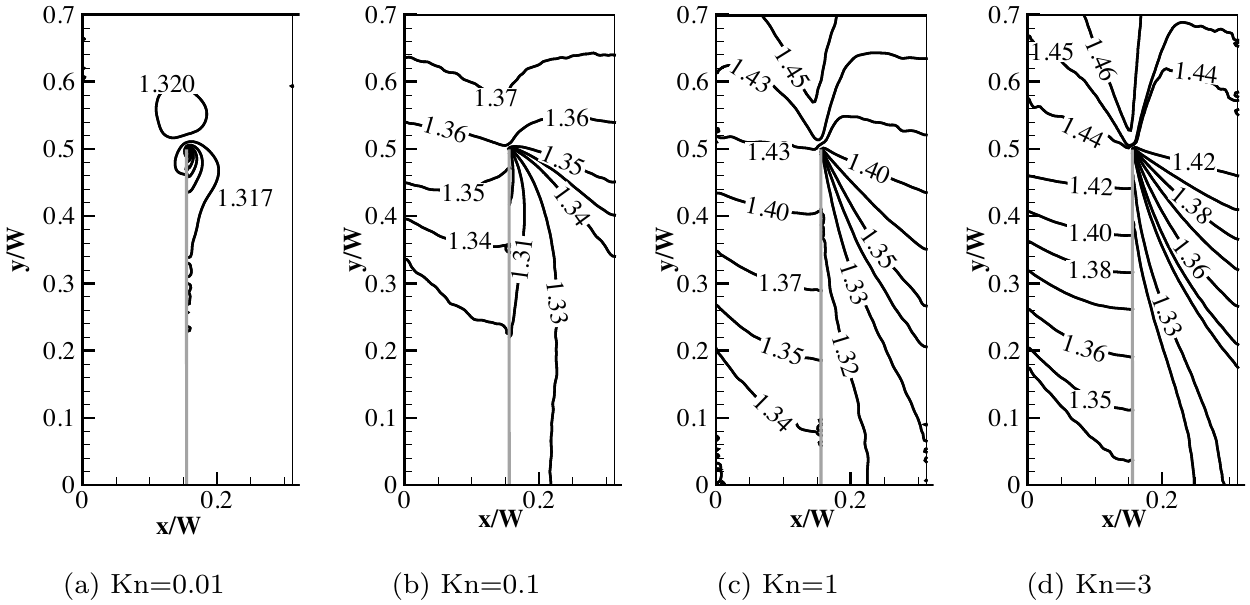}
	\end{center}	
	\caption{Isolines of pressure, $p/p_0$, for various Kn ($H/W=0.5$ and $H/L=1.6$).}\label{fig:iso_p}
\end{figure}

Before delving further into the dependence of the force on Kn, we show isolines of temperature and pressure as well as streamlines in figures \ref{fig:iso_T}, \ref{fig:iso_p} and \ref{fig:stream} for $H/W{=}0.5$ and $H/L{=}1.6$ at various Kn. In figure \ref{fig:iso_T} we observe that the pattern of temperature isolines at the tip of the vane is not radically different for the different Knudsen numbers. It is apparent, however, that the temperature gradient at the tip of the vane becomes more and more pronounced for smaller Knudsen numbers. Of course one has to keep in mind that particularly for Kn{\textgreater}1 the phase space distribution function is very different from an isotropic Maxwell distribution described by the scalar quantities density and temperature, but has a strong direction dependence in the velocity. For all Knudsen numbers shown this is also reflected in the temperature jump between a diffuse wall and the adjacent gas, which becomes larger with increasing mean free path.

\begin{figure}
	\begin{center}
		\includegraphics[scale=1]{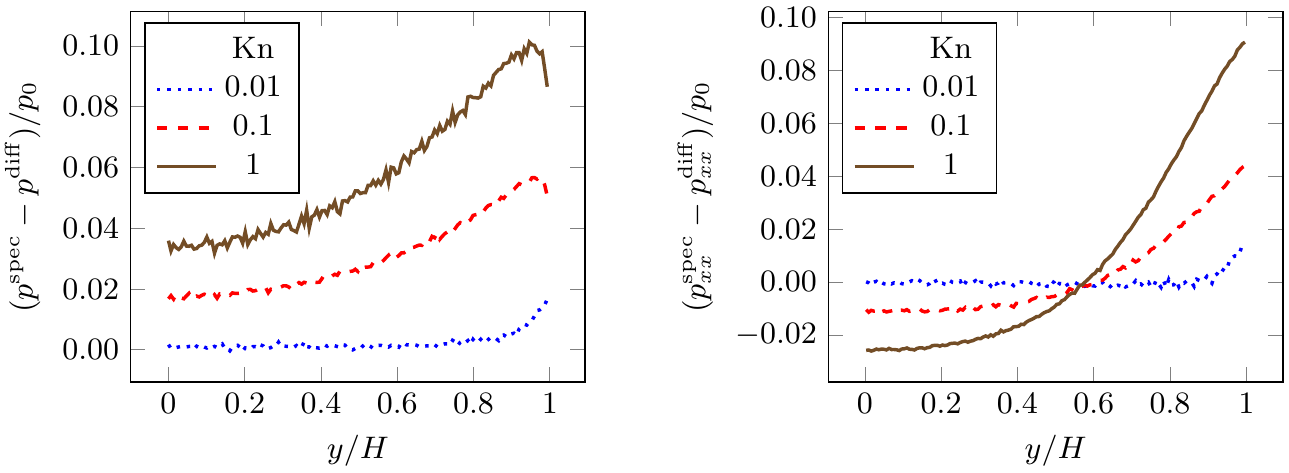}
	\end{center}
	\caption{Left: normalized pressure difference across the vane as function of distance from the centreline for Kn=0.01, 0.1, 1. Right: normalized difference in ${p}_{\mathit{xx}}$ across the vane (i.e. net force density in $x$-direction on the vane) as function of distance from the centreline ($H/W=0.5$ and $H/L=1.6$).}\label{fig:dF_vane}
\end{figure}

The pressure distribution, $p/p_0$, around the vane is shown in figure \ref{fig:iso_p}. At large Kn, the isobars follow roughly the same pattern as the isotherms. At $\mathrm{Kn}{\lesssim} 0.1$, however, some isobars originating from the tip of the vane curve back onto the diffuse side of the vane, and for lower Kn only the pressure in a small region extending a few mean free paths around the tip of the vane differs appreciably from its average value in the surroundings. This is an indication that for small Kn the main contribution to the net force is concentrated at the tip, extending only a few mean free paths along the vane. Note that the force on the vane is given by the integral of $p_{xx}$ along the vane. Due to the temperature gradient across the vane, the pressure tensor $p_{ij}$ is not isotropic, such that $p_{xx} {\neq} p_{yy}$. Therefore, $p$ is not the only contribution to the force on the vane. This is investigated in figure \ref{fig:dF_vane}, where the difference in $p$ (left) as well as $p_{xx}$ (right) between both sides of the vane is plotted as a function of the position along the vane. As can be seen, even for Kn as small as 0.1, the net force density does not vanish at the centre of the vane, and in this particular geometry also changes its sign, such that the gas pushes towards the left at the centre of the vane and to the right towards its edges. Only for smaller Kn, to a good approximation the net force density indeed vanishes everywhere except at the tip. Since the Knudsen number, $\mathrm{Kn}{=}\ell/W$, is based on the channel width, we have $\ell/H{=}0.02$ at Kn=0.01 and $W{=}2H$, such that the width of the edge zone is roughly $5\ell$ in this case.

\begin{figure}
	\begin{center}
		\includegraphics[scale=1]{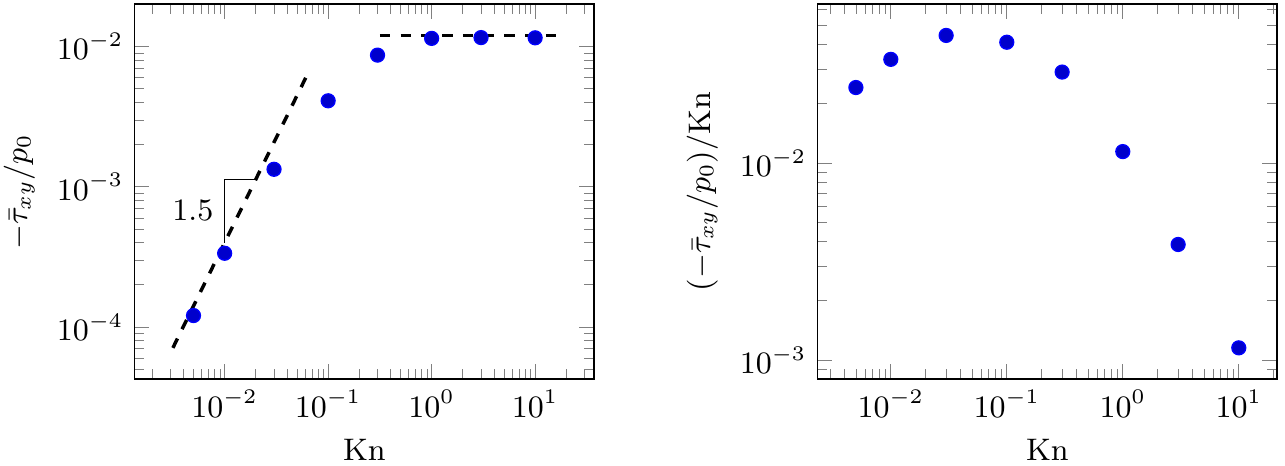}
	\end{center}
	\caption{Left: Normalized force density on the channel sidewall, $\bar{\tau}_{xy}/p_0$, as function of Kn at $H/W=0.5$ and $H/L=1.6$. Right: same data as left but using the pressure scale $\hat{p}_0=p_0\,\mathrm{Kn}$ for normalization, as discussed in the text.}\label{fig:force_Kn}
\end{figure}

In the left panel of figure \ref{fig:force_Kn} we plot the normalized mean shear stress, $-\bar{\tau}_{xy}/p_0$, on the channel walls as a function of Kn. We remind the reader that due to momentum conservation the mean force per unit area on the vane is $\bar{\mathcal{F}}_x|_\mathrm{vane}=-(L/H) \bar{\tau}_{xy}|_\mathrm{side}$. Evidently, $\bar{\tau}_{xy}/p_0$ tends to a constant for large Kn, while for small Kn it scales as ${\sim}\mathrm{Kn}^{1.5}$. Since in a typical experiment Kn is varied by changing the pressure, $p_0 {\sim} \mathrm{Kn}^{-1}$, the force density on the vane thus scales as $\bar{\mathcal{F}}_x{\sim}\mathrm{Kn}^{0.5}$ for small Kn and ${\sim}\mathrm{Kn}^{-1}$ for large Kn, and becomes extremal for Kn between 0.01 and 0.1. This is illustrated in figure \ref{fig:force_Kn} on the right, where we again use the reference pressure $\hat{p}_0=p_0\,\mathrm{Kn}$. Note that this scaling of the force with Kn agrees with results obtained for purely diffuse plates with sides held at different temperatures \citep{Taguchi_2012, Selden_2009a}.

Before turning to the performance of the setup as a pump, we give a qualitative explanation for the magnitude of the force as a function of Kn in the next section. 

%%%%%%%%%%%%%%%%%%%%%%%%%%%%%%%%%%%%%%%%%%%%%%%%%%%%%%%%
\section{Scaling analysis}\label{sec:normalStress}

\begin{figure}
\begin{center}
	\includegraphics[scale=1]{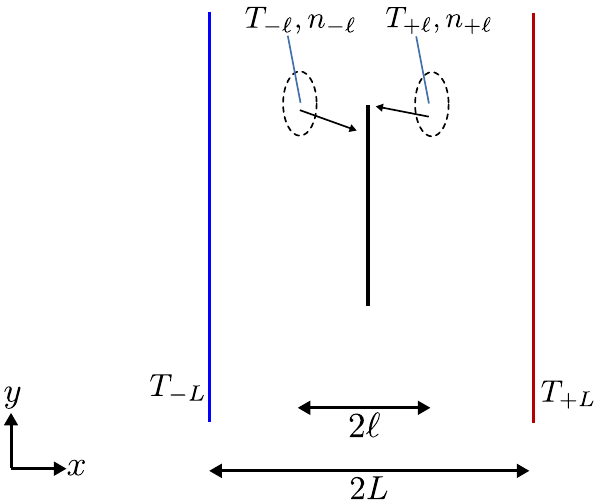}
\end{center}
\caption{Sketch of a plate inserted in the space between two plates at temperatures ${T}_{-L}$ and ${T}_{+L}$ respectively. For ${T}_{-L}<{T}_{+L}$, due to the net heat flux towards the left, molecules impinging onto the plate from the right carry a larger momentum than molecules from the left, leading to a net force on the plate towards the left normal to the face of the plate. If the plate is stationary, a corresponding gas flow around the plate edges towards the right is induced (from cold to hot).}\label{fig:sketch_planar}
\end{figure}

In figure \ref{fig:iso_T} we saw that the specular-diffuse plate introduces a strongly asymmetric temperature profile around the edge of the vane which leads to a net normal stress on the vane. To discuss such normal stresses on an object in a temperature gradient in rarefied gases, we consider a simplified geometry shown in figure \ref{fig:sketch_planar}, where a small plate is located midway between two plates of infinite extent at different temperatures. Although not strictly necessary for the argument, consider for the moment the case where the small plate's temperature is midway between the temperatures of the two plates and that the plate's sides are either purely diffusely or specularly reflecting. Due to the temperature gradient over the plate, on average the right side is hit by slightly faster molecules than the left side. The incoming gas molecules thus impart a net momentum in $-x$-direction onto the plate from the hotter to the colder side. Similarly, unless both sides of the plate are diffusely reflecting at the same temperature, on average the reflected particles contribute a normal force on the plate that is again larger on the hotter side. By momentum conservation an opposite force of equal magnitude is imparted onto the reflected gas molecules by the plate. Note that without the small plate, the gas phase would be at rest.

The magnitude of this force can be estimated as follows, loosely following an argument by \citet{Einstein_1924}. Molecules arriving from the right and left side of the plate will have a characteristic molecular velocity along the $x$-axis $\bar{c}_{\pm \ell}=\sqrt{2k_B T_{\pm\ell}/m}$ depending on the temperature at the position one mean free path away from the plate, as indicated by the corresponding subscript and sketched in figure \ref{fig:sketch_planar}. For situations where the gas velocity is much smaller than the characteristic molecular velocity, i.e. low Mach number, the molecular flux of particles from the right towards the edge of the plate is thus approximately $\sfrac{1}{2}\,n_{+\ell} \bar{c}_{+\ell}$, where $n_{+\ell}$ is the number density one mean free path to the right of the edge, and we have assumed that half of the particles originating from that position move to the left with a characteristic velocity $\bar{c}_{+ \ell}$ in $x$-direction. If no plate were present, this flux must be compensated by a corresponding flux $\sfrac{1}{2}\,{n}_{-\ell}{ \bar{c}}_{-\ell}$ of molecules from the left, and we assume this relation to hold approximately also at the edge of the plate, at least within a region of size $\ell$. Therefore, ${n}_{+\ell}{ \bar{c}}_{+\ell}{\approx}{n}_{-\ell}{ \bar{c}}_{-\ell}$, which we abbreviate as $n \bar{c}$. The corresponding flux of $x$-momentum carried in $x$-direction by these particles is thus $\left(\sfrac{1}{2}\,{n}_{\pm \ell}{\bar{c}}_{\pm \ell}\right)\left( m\bar{c}_{\pm l}\right) \approx\sfrac{1}{2}\,n\bar{c}m\bar{c}_{\pm l}$. Since they hit opposite sides of the plate they impart a net force per unit area of
\begin{equation}\label{eq:ForceEinstein}
\mathcal{F}_\mathrm{E} \approx \sfrac{1}{2}\, nm \bar{c} \left( \bar{c}_- - \bar{c}_{+} \right)
	\approx -nm\bar{c}\ell{{\partial}}_{x} \bar{c}
	\approx -k_Bn\ell \partial_x T 
	\approx -(p/T) \ell \partial_x T
\end{equation}
on the plate in $x$-direction, where $\bar{c}=\sqrt{2k_B T/m}$ was used to convert between the molecular velocity scale and the temperature. We stress that the derivative should be regarded as a shorthand for the normalized difference of properties approximately one mean free path to the left and right of the plate.

Apart from a factor of 1/2, equation (\ref{eq:ForceEinstein}) agrees with Einstein's 1924 estimate. Of course the total force on the plate is governed by both the momentum flux of particles impinging on and reflected from the surface. However, when both sides of the plate reflect diffusely at the same temperature the contribution of the reflected particles on both sides of the plate cancel (since the particle flux is equal) and for a purely specularly reflecting plate the reflected particles contribute with an equal magnitude as the impinging ones, since the normal component of the velocity of the impinging particles is reversed.

We thus stress that irrespective of the boundary condition at the wall there is a normal force on the plate as long as there is a temperature gradient normal to the plate. The boundary condition on the plate itself may be the origin of this temperature gradient, for example when the accommodation coefficient or wall temperature differs on opposite sides of the plate. However, the temperature gradient is usually also affected by the boundary conditions far away from the plate. Furthermore, the accommodation coefficient influences the magnitude of the force via the momentum transferred by the recoil of the reflected molecules.

Equation (\ref{eq:ForceEinstein}) is an estimate of the force density within a region extending a distance of $\ell$ away from the edge of the plate. To find the force on the plate we consider three different regimes depending on the Knudsen number. To begin, we start in a regime where the density is so low that the mean free path $\ell$ becomes comparable to the size of the container the plate is embedded in. In this case the thermal velocity of the gas molecules is essentially governed by the temperature of the last diffuse wall (container boundary or plate) from which these molecules scatter. An appropriate estimate of the force density is obtained by replacing $\ell \partial_x T$ in eq. (\ref{eq:ForceEinstein}) with the temperature difference $\Delta T$ between the plate and the corresponding wall. The force is obtained by multiplying the force density with the plate area $d^2$, such that $F {\sim} {-} d^2 nk_B \Delta T$, which in particular scales ${\sim} n$ for small gas densities. Since the product $n\ell$ is a constant, this means that $F {\sim} {\mathrm{Kn}}^{-1}$ in this regime. This can be considered as the free molecular regime discussed in section \ref{sec:collisionless}.

Next, consider a regime where $\ell$ is larger than the plate dimension $d$ but smaller than the container dimension. In this case the temperature field in the gas is governed by continuum-scale convective-diffusive heat transport, apart from a region a distance ${\sim}\ell$ away from the walls, and (\ref{eq:ForceEinstein}) is an adequate approximation on the entire plate surface. Thus $F {\sim} {-}d^2 k_B n \ell \partial_x T$, which, since $n\ell$ is constant, is a regime where the force is approximately independent of the Knudsen number. This is essentially the regime considered for thermophoresis of small particles  \citep{Einstein_1924, Waldmann_1959, Bakanov_1960}.

Finally, when the mean free path $\ell$ becomes much smaller than the plate dimensions, we have to consider the fact that (\ref{eq:ForceEinstein}) is only valid at the edge of the plate. Unless the external setup is such that the system works as a pump, the pressures on both sides of the plate are equal, and viscous or thermal stresses in the gas are negligible away from the edges of the plate. Since only a region of the extension of the mean free path around the edge contributes, the force on the plate scales as $F \sim -dk_B n \ell^2 \partial_x T$. In this regime, the force is thus expected to scale with $\ell \partial_x T$. This is the regime considered for radiometer forces by Einstein. For constant $\partial_{x}T$ the force scales like $F \sim n \ell^2 \sim \mathrm{Kn}$ in this regime. However, while (\ref{eq:ForceEinstein}) can be rigorously corroborated for large and intermediate Kn as long as the molecular velocity distribution is dominated by boundary conditions far away from the plate, for low Kn the temperature profile may be strongly influenced by the presence of the plate. Similar to the scaling of thermal edge flow at the edge of a heated plate found by \citet{Sone_1997}, one often finds a dependence closer to $\ell \partial_x T {\sim} \mathrm{Kn}^{1/2}$ within a region of size $\ell$ around the tip in the radiometer case as well, and the corresponding force-scaling becomes $F {\sim} \mathrm{Kn}^{1/2}$, cf. \citet{Taguchi_2012}. In the present case, an analogous scaling should be expected due to the similarity of the temperature field. Note that a similar scaling was employed by \citet{Wang_2016} in the context of thermally induced channel flow at the tips of ratchet teeth protruding into a channel.

In total, from the physical picture sketched above it can be inferred that the force on the plate has a maximum at an intermediate Kn number and drops off ${\sim} \mathrm{Kn}^{0.5}$ and ${\sim}\mathrm{Kn}^{-1}$ for smaller and larger Kn, respectively. This is indeed what is observed in the simulation results shown in figure \ref{fig:force_Kn}, corroborating our reasoning.

%%%%%%%%%%%%%%%%%%%%%%%%%%%%%%%%%%%%%%%%%%%%%%%%%%%%%%%%
%\subsection*{Comparison to other types of thermally induced flow}

The qualitative derivation of (\ref{eq:ForceEinstein}) is analogous to the explanation given for thermal creep flow along a wall with a tangential temperature gradient \citep{Sone_2007, Aoki_1995}. Here, again molecules impinging onto the wall from a region of higher temperature impart a larger momentum than molecules originating from a colder region. In the case of a diffusely reflecting wall this leads to a net tangential force on the wall towards the colder region. Conversely, for a stationary wall a creep flow along the wall towards the warmer region is induced in the gas phase, described by the Maxwell slip boundary condition \citep{Sone_2007, Lockerby_2004}. In the present case of specular-diffuse vanes in a channel, since the corresponding force density is tangential to the boundary, it does not directly contribute to the net force on the vanes in channel direction, i.e. normal to the vanes. However, for a plate of finite thickness with a temperature gradient along its edge, this tangential force component can contribute directly to the force on the Crookes radiometer \citep{Hettner_1924, Scandurra_2007, Ventura_2013}. 

\section{Pumping performance}\label{sec:MassFlux}

\begin{figure}
	\begin{center}
		\includegraphics[scale=1]{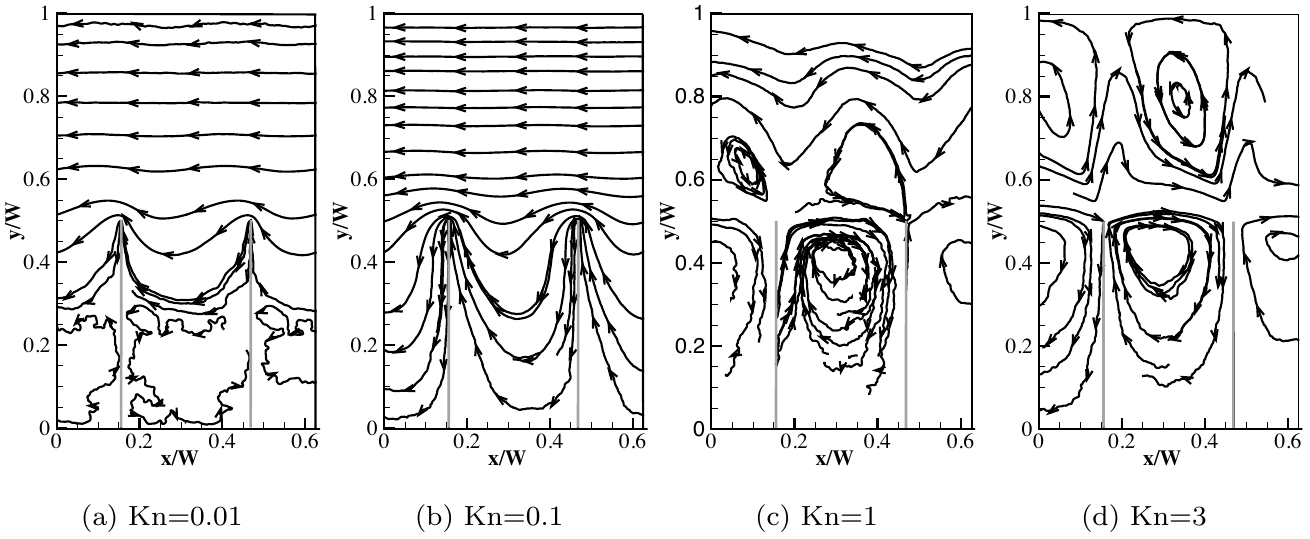}
	\end{center}
	\caption{Streamline patterns at different Kn (for $H/W=0.5$ and  $H/L=1.6$).}\label{fig:stream}
\end{figure}

\begin{figure}
	\begin{center}
		\includegraphics[scale=1]{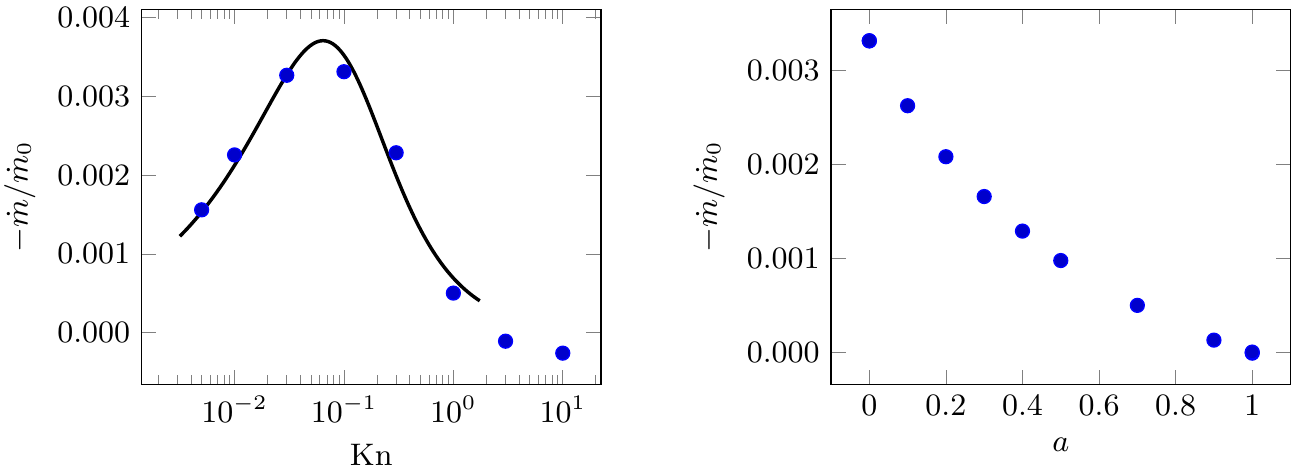}
	\end{center}
	\caption{Left: Normalized mass flow in the channel as a function of Kn. The solid line corresponds to a least squares fit of $g(\mathrm{Kn};a,b) {=} (a\,\mathrm{Kn}^{-0.5}+b\,\mathrm{Kn})^{-1}$ to the values for Kn${\leq} 1$. ($H/W=0.5$ and $H/L=1.6$). Right: Normalized mass flow in the channel as function of the accommodation coefficient $a$ (Kn=0.1, $H/W$=0.5 and $H/L$=1.6).}\label{fig:massFlow_Kn}

\end{figure}

Since the gas exerts a net force in $x$-direction on the vanes, an equal but opposite force acts on the reflected gas molecules due to the interaction. At finite Kn this leads to a net flow of the gas phase, as shown in figure \ref{fig:stream}. Since the net force on the vane is towards the right, a net flow towards the left is induced in the gas. For Kn up to 0.1 the tip of the vane acts similar to a localized volume force density in its vicinity, driving the gas towards the left such that the flow pattern resembles a Couette flow in the region between the vanes and the channel wall. As the mean free path becomes of the same order of magnitude of the channel dimension, the details of the flow field are expected to be significantly influenced by boundary conditions farther away and not solely by the local temperature field at the edge of the vane. Indeed, for Kn=1 the flow field can be seen to become qualitatively different from the Couette like flow for lower Kn, with a vortex emanating from the specular side of the vane reaching into the channel. This vortex grows with the mean free path and at Kn=3 eventually fills the whole region between the vanes and the sidewalls of the channel. This change in flow pattern has interesting consequences for the mass flow along the channel shown in figure \ref{fig:massFlow_Kn} (left side).  While for small Kn the flow is in negative $x$-direction, for large Kn a net flow in the opposite direction is observed.

The magnitude of the net mass flux in the continuum regime can be estimated on the basis of the discussion of the forces on the vanes and sidewalls in section \ref{sec:normalStress}. Within a region of the size of one mean free path around the edge, momentum is exchanged between the vane and gas. At low Kn the effect of this is similar to that of localized momentum sources at the edges of the vanes, while the Stokes equation governs the velocity within the rest of the domain, resulting in a velocity profile similar to Couette flow within the open section of the channel. The velocity is thus expected to scale as $\bar{u}{\sim} \tau h/\eta$, where $\tau$ is the shear rate at the side wall, $h$ is a length scale of the order of the distance between the vane and side walls, and $\eta$ is the mean viscosity. Since at moderate pressures the viscosity does not depend on pressure, the dependence of the velocity on Kn is the same as for the force on the vane and sidewall.

On the other extreme of very large Knudsen numbers, i.e. the collisionless regime, the gas velocity vanishes everywhere  \citep{Sone_2007}. For large but finite Kn numbers we can imagine the phase space distribution function, $f=f_\infty+\delta f$, to become altered slightly relative to the collisionless distribution, $f_{\infty}$, by few collisions between the gas molecules. Since the collision probability for each molecule is proportional to the gas density, the relative change in phase space distribution, $\delta f/f_{\infty}$, and hence the mean gas velocity are expected to be proportional to the gas density, i.e. $\bar{u}/c_0 \sim \mathrm{Kn}^{-1}$ in this regime.

Overall, the non-dimensionalized flow velocity and hence the non-dimensionalized mass flux are thus expected to scale ${\sim} \mathrm{Kn}^{0.5}$ and ${\sim} \mathrm{Kn}^{-1}$ for small and large Kn, respectively. In order to capture both limits, the function $g(\mathrm{Kn};a,b) {=} (a\,\mathrm{Kn}^{-0.5}+b\,\mathrm{Kn})^{-1}$ was used in a least-squares fit for the positive data points on the left panel in figure \ref{fig:massFlow_Kn} and plotted as a solid line. As can be seen, the function captures the dependence on Kn very well, corroborating our argument. However, we stress again that this can only be a crude approximation and not too much importance should be assigned to the quality of this fit, since as we have seen in particular for Kn{\textgreater}1 that the boundary conditions far away from the vane can lead to a qualitatively different flow field than at intermediate and small Kn. We briefly demonstrate this in  appendix \ref{sec:edgeVortex}, where the flow pattern and net mass flow at large Kn for vanes occupying different fractions $H/W$ of the channel are discussed. Additionally, we show that significantly larger flow rates can be obtained by reducing the vane size. As shown in the right panel of figure \ref{fig:variableHW}, in terms of an optimal geometry for a Knudsen pump, a vane of size $H/W$=0.1 performs at a roughly 5 times higher flow rate at Kn=0.1 compared to the case $H/W$=0.5 discussed so far.

So far we have considered the left sides of the vanes to be perfectly specularly reflective. In order to assess the impact of a departure from this idealized condition on the pumping performance, we relax this assumption and consider a situation where the left sides are characterized by a Maxwell-type boundary condition \citep{Sone_2007} with accommodation coefficient $a$ between 0 (fully specular) and 1 (fully diffuse). For molecules reflected diffusely from these walls the temperature is set to $T_1$, as on the right side of the vanes. The normalized mass flow in this situation is plotted as a function of $a$ on the right panel of figure \ref{fig:massFlow_Kn}. It is evident that even for a partially specular wall substantial mass flows are observed. However, since our focus was expressly on the impact of walls of high specular reflectivity we will not pursue this further here.

The pumping performance of the idealized pump studied here can be compared to other configurations described in the literature that are either geometrically similar or use a similar pumping mechanism for generating an inhomogeneous temperature profile along a channel. Although a direct comparison is difficult due to the different simulation parameters used in the studies, in particular with respect to the specific geometry and applied temperature range, a qualitative comparison is feasible, assuming linearity of the pumping performance with the temperature difference applied. In the geometrically similar configuration studied by \citet{Taguchi_2015} the temperature profile in the channel is generated by applying different temperatures on both sides of the vanes, considering fully diffuse reflection. In this set-up, the mass flow is slightly larger than in the present configuration, but of the same order of magnitude. The better performance of this configuration most likely stems from the larger temperature gradients at the tips of the vanes due to directly prescribing the temperatures instead of indirectly generating a temperature profile via a specular boundary. \citet{Wang_2016} considered a channel with one flat, diffusely reflecting wall and an opposing surface structured as a ratchet with parts of each ratchet tooth reflecting diffusely and parts fully specularly, similarly to the vanes considered here. The temperature profile in the channel is generated by prescribing different temperatures at the opposing walls. The mass flow obtained with this configuration is of the same order of magnitude as the one observed here. In a similar configuration, \citet{Chen_2016} used two opposing ratchet surfaces and considered several different reflection properties at the walls. In the situation where the opposing walls are fully diffusely reflecting, the mass flow rates obtained are 1-2 orders of magnitude smaller than in the present case. However, when using alternating diffuse and purely specular segments on each ratchet tooth, the mass flow rates obtained are again of the same order of magnitude as the ones in the present set-up.

%%%%%%%%%%%%%%%%%%%%%%%%%%%%%%%%%%%%%%%%%%%%%%%%%%%%%%%%
\section{Conclusions}\label{sec:conclusions}

\citet{Maxwell_1879} proposed a simple model for the gas-surface interaction that is intermediate between diffuse reflection and specular reflection, with an accommodation coefficient specifying the fraction of molecules reflected diffusely while the rest is reflected specularly. Due to its simplicity this model is widely used both in simulations as well as for characterizing experimental data, see the reviews of \citet{Agrawal_2008} and \citet{Cao_2009}. Note that \citet{Knudsen_1930} interpreted his experimental results on radiometric forces by invoking a difference in accomodation coefficients on opposite sides of a vane. In this paper our focus was on the impact of walls with low  accommodation upon reflection of molecules, and we therefore considered the extreme cases of ideal diffuse and ideal specular reflection. Although most surfaces have accommodation coefficients not too far away from 1, we stress that this is not merely an academic exercise. Low values in the range of 0.1--0.4 have been obtained both experimentally and numerically for certain gas-surface combinations, opening the possibility of a customized patterning of surface reflectivities. 

Specular reflection is often considered to be detrimental to thermally induced rarefied gas flows. Here we have shown that a specularly reflecting surface does not only shape the temperature profile in an essential way, but also argue that it contributes directly to flow over edges where a normal temperature gradient occurs. In particular, we have studied a Knudsen pump inspired by the Crookes radiometer with vanes placed into an inhomogeneous temperature field. Contrary to the Crookes radiometer, where the temperature gradient comes about by selectively heating one side of each vane (thus giving opposite sides different temperatures), in the present case the temperature field is shaped in an essential way by a specularly reflecting boundary on one side of the vane. The origin of the observed forces on the vanes and their dependence on the Knudsen number was discussed based on a simple argument by Einstein. As was pointed out already by \citet{Aoki_1995}, at low Kn numbers the corresponding flow across the vanes is essentially of the same kind as thermal creep flow along a wall with a tangential temperature gradient or thermal edge flow at a vane with a different temperature than its surroundings. As such, at low Kn it can be interpreted as a pure boundary effect, needed to reconcile the continuum field equations with the discontinuous phase space density at a wall. For the similar flow across the vane of a Crookes radiometer \citet{Taguchi_2012, Taguchi_2015} coined the term radiometric flow.

Utilizing specular walls in a Knudsen pump offers several advantages. In the present geometry, apart from shaping the temperature profile in an essential way, the specular walls do not themselves partake in the exchange of energy between the hot and cold sections of the domain. This may be advantageous for the efficiency of the pump. Furthermore, due to the large velocity slip at walls with low accommodation coefficients, the inclusion of such walls into a Knudsen pump potentially results in larger flow velocities due to reduced viscous dissipation in the system. However, in the current setup with vanes normal to the flow direction this will be a small effect, but should boost the performance in systems with slanted sidewalls such as ratchets \citep{Donkov_2011, Chen_2016, Wang_2016}.

%%%%%%%%%%%%%%%%%%%%% Acknowledgments %%%%%%%%%%%%%%%%%%%%%
\begin{acknowledgments}
	All numerical calculations were done on the Lichtenberg high performance computer at the TU Darmstadt. E.R. is indebted to the DAAD (Deutscher Akademischer Aus\-tausch\-dienst) for financial support of his visit to Darmstadt.  The authors would like to thank Tom Scanlon from the University of Strathclyde for providing access to the dsmcFoamStrath solver. Finally, we express our gratitude to Craig White from the University of Glasgow for sharing his expertise in dsmcFoamStrath.
\end{acknowledgments}

%%%%%%%%%%%%%%%%%%%%%%%% Appendix %%%%%%%%%%%%%%%%%%%%%%%%

%\balancecolsandclearpage
%\clearpage

\appendix
%\section*{Appendix}
%%%%%%%%%%%%%%%%%%%%%%%%%%%%%%%%%%%%%%%%%%%%%%%%%%%%%%%%
\section{Flow field at large \texorpdfstring{K\MakeLowercase{n}}{Kn} and variation of H/W}\label{sec:edgeVortex}

\begin{figure}
	\begin{center}
		\includegraphics[scale=1]{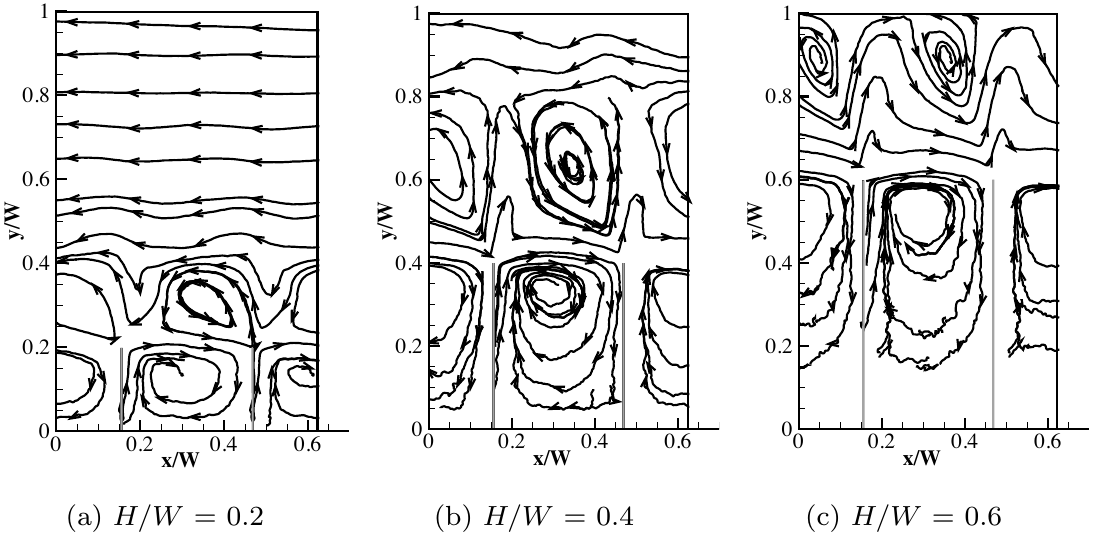}
	\end{center}
	\caption{Velocity field at Kn = 3 and $W/L$=3.2 for various $H/W$.}\label{fig:stream_Kn3}
\end{figure}

\begin{figure}
	\begin{center}
		\includegraphics[scale=1]{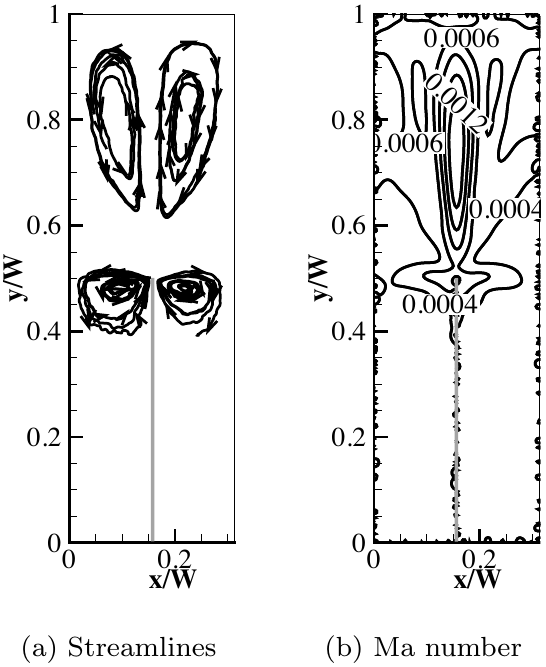}
	\end{center}
	\caption{Fully diffuse vane array at Kn = 3, $H/W$=0.5 and $W/L$=3.2. The temperature ratio between sidewall and vanes is again $T_2/T_1=2$. Left: Streamlines, right: Isolines of local Mach number.}\label{fig:fullyDiffuseKn3}
\end{figure}

In section \ref{sec:MassFlux} we observed that for large Knudsen numbers the net mass flow direction can be opposite to the flow direction at small Kn. We identified the appearance of edge-vortices as the cause for this flow reversal. In order to further investigate the influence of the edge-vortex on the flow, we plot streamlines at Kn=3 for various values of $H/W$ in figure \ref{fig:stream_Kn3}. Note that in each case the vortex occupies roughly equal amounts of the channel, such that it influences the net flow strongest for small gaps between vane and channel wall. Indeed, at Kn=3, the net flow is in opposite direction to the direction of force on the sidewalls only for $H/W {\gtrsim} 0.4$.

Note that even though the vortex appears on the specular side, it is not a feature exclusive to this boundary condition. When both sides of the vane have the same temperature and are fully diffuse, a symmetric flow pattern with two counter-rotating vortices appears above the vane, see figure \ref{fig:fullyDiffuseKn3}. Additionally, two smaller vortices appear on both sides of the vane. Note that at the edge the maximum velocity is parallel to the vane in the main vortices, while it is perpendicular to the vane in the two smaller ones.

\begin{figure}
	\begin{center}
		\includegraphics[scale=1]{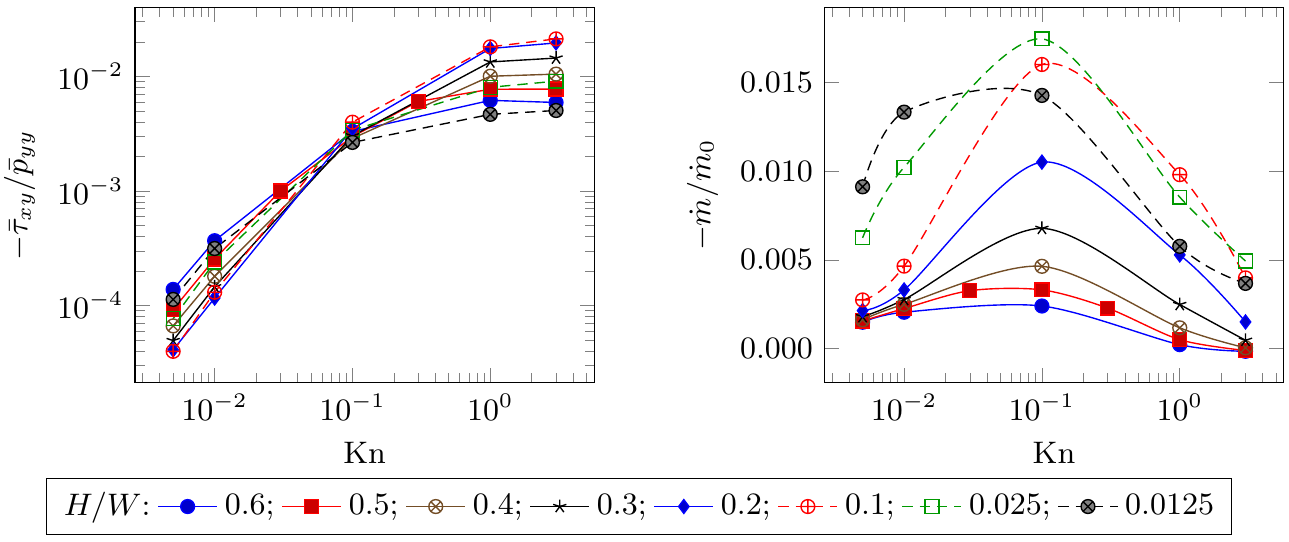}
	\end{center}
	\caption{Left: Normalized force density, $\bar{\tau}_{xy}/p_{yy}$, on the channel sidewall as function of Kn for variable $H/W$. Right: Normalized mass flow along the channel as function of Kn for variable $H/W$ (lines are merely guides for the eye).}\label{fig:variableHW}
\end{figure}

The forces and flow rates occurring for half-specular, half-diffuse vanes occupying different fractions of the channel are investigated in figure \ref{fig:variableHW}. The left panel in this figure shows the ratio between tangential and normal force on the sidewall for various vane heights as a function of Kn. As can be seen, the curves all follow the general pattern observed in figure \ref{fig:force_Kn}, although the magnitudes may vary. Similarly, the normalized mass flow shown in the right part of figure \ref{fig:variableHW} shows a similar dependence on Kn as in figure \ref{fig:massFlow_Kn} for all values of $H/W$. Note, however, that for smaller vanes a significantly larger mass flow can be reached than for the base case $H/W$=0.5, considered in the main part of the paper. This is most likely due to the larger space between vanes and sidewalls. Also note that a mass flow in positive $x$-direction is only observed for large Kn numbers and large vanes, when the edge vortex discussed above in figure \ref{fig:stream_Kn3} occupies the whole width of the channel.

\section{Tabulated numerical results}\label{sec:numRes}
In table \ref{tab:numRes} we list numerical results for area-averaged and normalized forces and heat flows on the sidewall, ($\bar{\tau}_{xy}$, $\bar{p}_{yy}$, $\bar{q}_y$), and mass flow along the channel, $\dot{m}$. The relative error, $\sigma(X)/|X|$, of a quantity $X$ is estimated via the standard deviation from (at least three) consecutive runs. The heat flux at position $\mathbf{r}$ was defined as $\mathbf{q}\left(\mathbf{r}\right)=\left(m/{2}\right){\int}{\left(\mathbf{c}-\mathbf{u}\right)}{\left(\mathbf{c}-\mathbf{u}\right)}^{2}f\left(r,c\right){d}^{3}c$. All results are for $W/L$=3.2. We use the abbreviated scientific notation $a_n \equiv a\cdot 10^n$ for $a\in\mathbb{R}$ and $n\in\mathbb{Z}$.

%%%%%%%%%%%%%%%%%%%%%%% Bibliography %%%%%%%%%%%%%%%%%%%%%%%
%\bibliography{litVane}

\begin{thebibliography}{55}%
\makeatletter
\providecommand \@ifxundefined [1]{%
 \@ifx{#1\undefined}
}%
\providecommand \@ifnum [1]{%
 \ifnum #1\expandafter \@firstoftwo
 \else \expandafter \@secondoftwo
 \fi
}%
\providecommand \@ifx [1]{%
 \ifx #1\expandafter \@firstoftwo
 \else \expandafter \@secondoftwo
 \fi
}%
\providecommand \natexlab [1]{#1}%
\providecommand \enquote  [1]{``#1''}%
\providecommand \bibnamefont  [1]{#1}%
\providecommand \bibfnamefont [1]{#1}%
\providecommand \citenamefont [1]{#1}%
\providecommand \href@noop [0]{\@secondoftwo}%
\providecommand \href [0]{\begingroup \@sanitize@url \@href}%
\providecommand \@href[1]{\@@startlink{#1}\@@href}%
\providecommand \@@href[1]{\endgroup#1\@@endlink}%
\providecommand \@sanitize@url [0]{\catcode `\\12\catcode `\$12\catcode
  `\&12\catcode `\#12\catcode `\^12\catcode `\_12\catcode `\%12\relax}%
\providecommand \@@startlink[1]{}%
\providecommand \@@endlink[0]{}%
\providecommand \url  [0]{\begingroup\@sanitize@url \@url }%
\providecommand \@url [1]{\endgroup\@href {#1}{\urlprefix }}%
\providecommand \urlprefix  [0]{URL }%
\providecommand \Eprint [0]{\href }%
\providecommand \doibase [0]{http://dx.doi.org/}%
\providecommand \selectlanguage [0]{\@gobble}%
\providecommand \bibinfo  [0]{\@secondoftwo}%
\providecommand \bibfield  [0]{\@secondoftwo}%
\providecommand \translation [1]{[#1]}%
\providecommand \BibitemOpen [0]{}%
\providecommand \bibitemStop [0]{}%
\providecommand \bibitemNoStop [0]{.\EOS\space}%
\providecommand \EOS [0]{\spacefactor3000\relax}%
\providecommand \BibitemShut  [1]{\csname bibitem#1\endcsname}%
\let\auto@bib@innerbib\@empty
%</preamble>
\bibitem [{\citenamefont {Struchtrup}(2005)}]{Struchtrup_2005}%
  \BibitemOpen
  \bibfield  {author} {\bibinfo {author} {\bibfnamefont {Henning}\ \bibnamefont
  {Struchtrup}},\ }\href@noop {} {\emph {\bibinfo {title} {Macroscopic
  transport equations for rarefied gas flows}}}\ (\bibinfo  {publisher}
  {Springer},\ \bibinfo {year} {2005})\BibitemShut {NoStop}%
\bibitem [{\citenamefont {{Sone}}(2007)}]{Sone_2007}%
  \BibitemOpen
  \bibfield  {author} {\bibinfo {author} {\bibfnamefont {Y.}~\bibnamefont
  {{Sone}}},\ }\href@noop {} {\emph {\bibinfo {title} {Molecular Gas Dynamics:
  Theory, Techniques, and Applications}}}\ (\bibinfo  {publisher}
  {Birkh\"auser},\ \bibinfo {year} {2007})\BibitemShut {NoStop}%
\bibitem [{\citenamefont {Landau}\ and\ \citenamefont
  {Lifschitz}(1983)}]{Landau_1983_X}%
  \BibitemOpen
  \bibfield  {author} {\bibinfo {author} {\bibfnamefont {Lev~Davidovich}\
  \bibnamefont {Landau}}\ and\ \bibinfo {author} {\bibfnamefont
  {EM}~\bibnamefont {Lifschitz}},\ }\href@noop {} {\emph {\bibinfo {title}
  {Physikalische Kinetik}}}\ (\bibinfo  {publisher} {Akademie-Verlag},\
  \bibinfo {year} {1983})\BibitemShut {NoStop}%
\bibitem [{\citenamefont {Reif}(1965)}]{Reif_1965}%
  \BibitemOpen
  \bibfield  {author} {\bibinfo {author} {\bibfnamefont {Frederick}\
  \bibnamefont {Reif}},\ }\href@noop {} {\emph {\bibinfo {title} {Fundamentals
  of statistical and thermal physics}}}\ (\bibinfo  {publisher} {McGraw-Hill},\
  \bibinfo {year} {1965})\BibitemShut {NoStop}%
\bibitem [{\citenamefont {Tyndall}(1870)}]{Tyndall_1870}%
  \BibitemOpen
  \bibfield  {author} {\bibinfo {author} {\bibfnamefont {John}\ \bibnamefont
  {Tyndall}},\ }\bibfield  {title} {\enquote {\bibinfo {title} {On the action
  of rays of high refrangibility upon gaseous matter},}\ }\href {\doibase
  10.1098/rstl.1870.0019} {\bibfield  {journal} {\bibinfo  {journal} {Philos.\
  Trans.\ R.\ Soc.\ London}\ }\textbf {\bibinfo {volume} {160}},\ \bibinfo
  {pages} {333--365} (\bibinfo {year} {1870})}\BibitemShut {NoStop}%
\bibitem [{\citenamefont {Davis}\ and\ \citenamefont
  {Schweiger}(2002)}]{Davis_2002}%
  \BibitemOpen
  \bibfield  {author} {\bibinfo {author} {\bibfnamefont {E~James}\ \bibnamefont
  {Davis}}\ and\ \bibinfo {author} {\bibfnamefont {Gustav}\ \bibnamefont
  {Schweiger}},\ }\href@noop {} {\emph {\bibinfo {title} {The airborne
  microparticle: its physics, chemistry, optics, and transport phenomena}}}\
  (\bibinfo  {publisher} {Springer},\ \bibinfo {year} {2002})\BibitemShut
  {NoStop}%
\bibitem [{\citenamefont {Crookes}(1876)}]{Crookes_1876}%
  \BibitemOpen
  \bibfield  {author} {\bibinfo {author} {\bibfnamefont {William}\ \bibnamefont
  {Crookes}},\ }\href@noop {} {\enquote {\bibinfo {title} {Improvement in
  apparatus for indicating the intensity of radiation},}\ }\bibinfo
  {howpublished} {US patent 182172} (\bibinfo {year} {1876})\BibitemShut
  {NoStop}%
\bibitem [{\citenamefont {Reynolds}(1879)}]{Reynolds_1879}%
  \BibitemOpen
  \bibfield  {author} {\bibinfo {author} {\bibfnamefont {Osborne}\ \bibnamefont
  {Reynolds}},\ }\bibfield  {title} {\enquote {\bibinfo {title} {On certain
  dimensional properties of matter in the gaseous state.}}\ }\href {\doibase
  10.1098/rstl.1879.0078} {\bibfield  {journal} {\bibinfo  {journal} {Philos.\
  Trans.\ R.\ Soc.\ London}\ }\textbf {\bibinfo {volume} {170}},\ \bibinfo
  {pages} {727--845} (\bibinfo {year} {1879})}\BibitemShut {NoStop}%
\bibitem [{\citenamefont {Maxwell}(1879)}]{Maxwell_1879}%
  \BibitemOpen
  \bibfield  {author} {\bibinfo {author} {\bibfnamefont {J~Clerk}\ \bibnamefont
  {Maxwell}},\ }\bibfield  {title} {\enquote {\bibinfo {title} {On stresses in
  rarified gases arising from inequalities of temperature},}\ }\href {\doibase
  10.1098/rstl.1879.0067} {\bibfield  {journal} {\bibinfo  {journal} {Philos.
  Trans. R. Soc. London}\ }\textbf {\bibinfo {volume} {170}},\ \bibinfo {pages}
  {231--256} (\bibinfo {year} {1879})}\BibitemShut {NoStop}%
\bibitem [{\citenamefont {Knudsen}(1910)}]{Knudsen_1910}%
  \BibitemOpen
  \bibfield  {author} {\bibinfo {author} {\bibfnamefont {Martin}\ \bibnamefont
  {Knudsen}},\ }\bibfield  {title} {\enquote {\bibinfo {title} {Thermischer
  Molekulardruck der Gase in R\"ohren},}\ }\href {\doibase
  10.1002/andp.19103381618} {\bibfield  {journal} {\bibinfo  {journal} {Ann.
  Phys.}\ }\textbf {\bibinfo {volume} {338}},\ \bibinfo {pages} {1435--1448}
  (\bibinfo {year} {1910})}\BibitemShut {NoStop}%
\bibitem [{\citenamefont {Sone}\ \emph {et~al.}(1996)\citenamefont {Sone},
  \citenamefont {Waniguchi},\ and\ \citenamefont {Aoki}}]{Sone_1996}%
  \BibitemOpen
  \bibfield  {author} {\bibinfo {author} {\bibfnamefont {Yoshio}\ \bibnamefont
  {Sone}}, \bibinfo {author} {\bibfnamefont {Yorifumi}\ \bibnamefont
  {Waniguchi}}, \ and\ \bibinfo {author} {\bibfnamefont {Kazuo}\ \bibnamefont
  {Aoki}},\ }\bibfield  {title} {\enquote {\bibinfo {title} {One-way flow of a
  rarefied gas induced in a channel with a periodic temperature
  distribution},}\ }\href {\doibase 10.1063/1.869101} {\bibfield  {journal}
  {\bibinfo  {journal} {Phys.\ Fluids}\ }\textbf {\bibinfo {volume} {8}},\
  \bibinfo {pages} {2227--2235} (\bibinfo {year} {1996})}\BibitemShut {NoStop}%
\bibitem [{\citenamefont {Young}\ \emph {et~al.}(2003)\citenamefont {Young},
  \citenamefont {Han}, \citenamefont {Muntz}, \citenamefont {Shiflett},
  \citenamefont {Ketsdever},\ and\ \citenamefont {Green}}]{Young_2003}%
  \BibitemOpen
  \bibfield  {author} {\bibinfo {author} {\bibfnamefont {Marcus}\ \bibnamefont
  {Young}}, \bibinfo {author} {\bibfnamefont {Yen~Lin}\ \bibnamefont {Han}},
  \bibinfo {author} {\bibfnamefont {EP}~\bibnamefont {Muntz}}, \bibinfo
  {author} {\bibfnamefont {G}~\bibnamefont {Shiflett}}, \bibinfo {author}
  {\bibfnamefont {Andrew}\ \bibnamefont {Ketsdever}}, \ and\ \bibinfo {author}
  {\bibfnamefont {Amanda}\ \bibnamefont {Green}},\ }\bibfield  {title}
  {\enquote {\bibinfo {title} {Thermal transpiration in microsphere
  membranes},}\ }in\ \href {\doibase 10.1063/1.1581617} {\emph {\bibinfo
  {booktitle} {AIP Conference Proceedings}}}\ (\bibinfo {organization} {IOP
  Institute of Physics Publishing LTD},\ \bibinfo {year} {2003})\ pp.\ \bibinfo
  {pages} {743--751}\BibitemShut {NoStop}%
\bibitem [{\citenamefont {An}\ \emph {et~al.}(2014)\citenamefont {An},
  \citenamefont {Gupta},\ and\ \citenamefont {Gianchandani}}]{An_2014}%
  \BibitemOpen
  \bibfield  {author} {\bibinfo {author} {\bibfnamefont {Seungdo}\ \bibnamefont
  {An}}, \bibinfo {author} {\bibfnamefont {Neeraj~K}\ \bibnamefont {Gupta}}, \
  and\ \bibinfo {author} {\bibfnamefont {Yogesh~B}\ \bibnamefont
  {Gianchandani}},\ }\bibfield  {title} {\enquote {\bibinfo {title} {A
  Si-micromachined 162-stage two-part knudsen pump for on-chip vacuum},}\
  }\href {\doibase 10.1109/JMEMS.2013.2281316} {\bibfield  {journal} {\bibinfo
  {journal} {J. Microelectromech. Syst.}\ }\textbf {\bibinfo {volume} {23}},\
  \bibinfo {pages} {406--416} (\bibinfo {year} {2014})}\BibitemShut {NoStop}%
\bibitem [{\citenamefont {Sugimoto}\ and\ \citenamefont
  {Sone}(2005)}]{Sugimoto_2005}%
  \BibitemOpen
  \bibfield  {author} {\bibinfo {author} {\bibfnamefont {Hiroshi}\ \bibnamefont
  {Sugimoto}}\ and\ \bibinfo {author} {\bibfnamefont {Yoshio}\ \bibnamefont
  {Sone}},\ }\bibfield  {title} {\enquote {\bibinfo {title} {Vacuum pump
  without a moving part driven by thermal edge flow},}\ }in\ \href@noop {}
  {\emph {\bibinfo {booktitle} {Rarefied Gas Dynamics: 24th International
  Symposium on Rarefied Gas Dynamics RGD 24}}},\ Vol.\ \bibinfo {volume} {762}\
  (\bibinfo {year} {2005})\ pp.\ \bibinfo {pages} {168--173}\BibitemShut
  {NoStop}%
\bibitem [{\citenamefont {Donkov}\ \emph {et~al.}(2011)\citenamefont {Donkov},
  \citenamefont {Tiwari}, \citenamefont {Liang}, \citenamefont {Hardt},
  \citenamefont {Klar},\ and\ \citenamefont {Ye}}]{Donkov_2011}%
  \BibitemOpen
  \bibfield  {author} {\bibinfo {author} {\bibfnamefont {Alexander~A.}\
  \bibnamefont {Donkov}}, \bibinfo {author} {\bibfnamefont {Sudarshan}\
  \bibnamefont {Tiwari}}, \bibinfo {author} {\bibfnamefont {Tengfei}\
  \bibnamefont {Liang}}, \bibinfo {author} {\bibfnamefont {Steffen}\
  \bibnamefont {Hardt}}, \bibinfo {author} {\bibfnamefont {Axel}\ \bibnamefont
  {Klar}}, \ and\ \bibinfo {author} {\bibfnamefont {Wenjing}\ \bibnamefont
  {Ye}},\ }\bibfield  {title} {\enquote {\bibinfo {title} {Momentum and mass
  fluxes in a gas confined between periodically structured surfaces at
  different temperatures},}\ }\href {\doibase 10.1103/PhysRevE.84.016304}
  {\bibfield  {journal} {\bibinfo  {journal} {Phys.\ Rev.\ E}\ }\textbf
  {\bibinfo {volume} {84}},\ \bibinfo {pages} {016304} (\bibinfo {year}
  {2011})}\BibitemShut {NoStop}%
\bibitem [{\citenamefont {W{\"u}rger}(2011)}]{Wurger_2011}%
  \BibitemOpen
  \bibfield  {author} {\bibinfo {author} {\bibfnamefont {Alois}\ \bibnamefont
  {W{\"u}rger}},\ }\bibfield  {title} {\enquote {\bibinfo {title} {Leidenfrost
  gas ratchets driven by thermal creep},}\ }\href {\doibase
  10.1103/PhysRevLett.107.164502} {\bibfield  {journal} {\bibinfo  {journal}
  {Phys.\ Rev.\ Lett.}\ }\textbf {\bibinfo {volume} {107}},\ \bibinfo {pages}
  {164502} (\bibinfo {year} {2011})}\BibitemShut {NoStop}%
\bibitem [{\citenamefont {Chen}\ \emph {et~al.}(2014)\citenamefont {Chen},
  \citenamefont {Baldas},\ and\ \citenamefont {Colin}}]{Chen_2014}%
  \BibitemOpen
  \bibfield  {author} {\bibinfo {author} {\bibfnamefont {J}~\bibnamefont
  {Chen}}, \bibinfo {author} {\bibfnamefont {L}~\bibnamefont {Baldas}}, \ and\
  \bibinfo {author} {\bibfnamefont {S}~\bibnamefont {Colin}},\ }\bibfield
  {title} {\enquote {\bibinfo {title} {Numerical study of thermal creep flow
  between two ratchet surfaces},}\ }\href {\doibase
  10.1016/j.vacuum.2014.05.013} {\bibfield  {journal} {\bibinfo  {journal}
  {Vacuum}\ }\textbf {\bibinfo {volume} {109}},\ \bibinfo {pages} {294--301}
  (\bibinfo {year} {2014})}\BibitemShut {NoStop}%
\bibitem [{\citenamefont {Chen}\ \emph {et~al.}(2016)\citenamefont {Chen},
  \citenamefont {Stefanov}, \citenamefont {Baldas},\ and\ \citenamefont
  {Colin}}]{Chen_2016}%
  \BibitemOpen
  \bibfield  {author} {\bibinfo {author} {\bibfnamefont {Jie}\ \bibnamefont
  {Chen}}, \bibinfo {author} {\bibfnamefont {Stefan~K}\ \bibnamefont
  {Stefanov}}, \bibinfo {author} {\bibfnamefont {Lucien}\ \bibnamefont
  {Baldas}}, \ and\ \bibinfo {author} {\bibfnamefont {St{\'e}phane}\
  \bibnamefont {Colin}},\ }\bibfield  {title} {\enquote {\bibinfo {title}
  {Analysis of flow induced by temperature fields in ratchet-like microchannels
  by direct simulation Monte Carlo},}\ }\href {\doibase
  10.1016/j.ijheatmasstransfer.2016.04.023} {\bibfield  {journal} {\bibinfo
  {journal} {Int. J. Heat Mass Transfer}\ }\textbf {\bibinfo {volume} {99}},\
  \bibinfo {pages} {672--680} (\bibinfo {year} {2016})}\BibitemShut {NoStop}%
\bibitem [{\citenamefont {Wang}\ \emph {et~al.}(2016)\citenamefont {Wang},
  \citenamefont {Xu}, \citenamefont {Xu},\ and\ \citenamefont
  {Qian}}]{Wang_2016}%
  \BibitemOpen
  \bibfield  {author} {\bibinfo {author} {\bibfnamefont {Ruijie}\ \bibnamefont
  {Wang}}, \bibinfo {author} {\bibfnamefont {Xinpeng}\ \bibnamefont {Xu}},
  \bibinfo {author} {\bibfnamefont {Kun}\ \bibnamefont {Xu}}, \ and\ \bibinfo
  {author} {\bibfnamefont {Tiezheng}\ \bibnamefont {Qian}},\ }\bibfield
  {title} {\enquote {\bibinfo {title} {Onsager's cross coupling effects in gas
  flows confined to micro-channels},}\ }\href {\doibase
  10.1103/PhysRevFluids.1.044102} {\bibfield  {journal} {\bibinfo  {journal}
  {Phys. Rev. Fluids}\ }\textbf {\bibinfo {volume} {1}},\ \bibinfo {pages}
  {044102} (\bibinfo {year} {2016})}\BibitemShut {NoStop}%
\bibitem [{\citenamefont {Westphal}(1920)}]{Westphal_1920}%
  \BibitemOpen
  \bibfield  {author} {\bibinfo {author} {\bibfnamefont {Wilhelm~H}\
  \bibnamefont {Westphal}},\ }\bibfield  {title} {\enquote {\bibinfo {title}
  {Messungen am Radiometer II},}\ }\href {\doibase 10.1007/BF01332672}
  {\bibfield  {journal} {\bibinfo  {journal} {Z.\ Phys. A}\ }\textbf {\bibinfo
  {volume} {1}},\ \bibinfo {pages} {431--438} (\bibinfo {year}
  {1920})}\BibitemShut {NoStop}%
\bibitem [{\citenamefont {Einstein}(1924)}]{Einstein_1924}%
  \BibitemOpen
  \bibfield  {author} {\bibinfo {author} {\bibfnamefont {Albert}\ \bibnamefont
  {Einstein}},\ }\bibfield  {title} {\enquote {\bibinfo {title} {Zur Theorie
  der Radiometerkr{\"a}fte},}\ }\href {\doibase 10.1007/BF01328006} {\bibfield
  {journal} {\bibinfo  {journal} {Z.\ Phys.}\ }\textbf {\bibinfo {volume}
  {27}},\ \bibinfo {pages} {1--6} (\bibinfo {year} {1924})}\BibitemShut
  {NoStop}%
\bibitem [{\citenamefont {Hettner}(1924)}]{Hettner_1924}%
  \BibitemOpen
  \bibfield  {author} {\bibinfo {author} {\bibfnamefont {Gerhard}\ \bibnamefont
  {Hettner}},\ }\bibfield  {title} {\enquote {\bibinfo {title} {Zur Theorie des
  Radiometers},}\ }\href {\doibase 10.1007/BF01328008} {\bibfield  {journal}
  {\bibinfo  {journal} {Z.\ Phys.}\ }\textbf {\bibinfo {volume} {27}},\
  \bibinfo {pages} {12--22} (\bibinfo {year} {1924})}\BibitemShut {NoStop}%
\bibitem [{\citenamefont {Sexl}(1926)}]{Sexl_1926}%
  \BibitemOpen
  \bibfield  {author} {\bibinfo {author} {\bibfnamefont {Theodor}\ \bibnamefont
  {Sexl}},\ }\bibfield  {title} {\enquote {\bibinfo {title} {Zur Theorie der
  Radiometerwirkungen II},}\ }\href {\doibase 10.1002/andp.19263862403}
  {\bibfield  {journal} {\bibinfo  {journal} {Ann.\ Phys.}\ }\textbf {\bibinfo
  {volume} {386}},\ \bibinfo {pages} {800--806} (\bibinfo {year}
  {1926})}\BibitemShut {NoStop}%
\bibitem [{\citenamefont {Epstein}(1929)}]{Epstein_1929}%
  \BibitemOpen
  \bibfield  {author} {\bibinfo {author} {\bibfnamefont {Paul~S}\ \bibnamefont
  {Epstein}},\ }\bibfield  {title} {\enquote {\bibinfo {title} {Zur Theorie des
  Radiometers},}\ }\href {\doibase 10.1007/BF01338485} {\bibfield  {journal}
  {\bibinfo  {journal} {Z.\ Phys.}\ }\textbf {\bibinfo {volume} {54}},\
  \bibinfo {pages} {537--563} (\bibinfo {year} {1929})}\BibitemShut {NoStop}%
\bibitem [{\citenamefont {Knudsen}(1930)}]{Knudsen_1930}%
  \BibitemOpen
  \bibfield  {author} {\bibinfo {author} {\bibfnamefont {Martin}\ \bibnamefont
  {Knudsen}},\ }\bibfield  {title} {\enquote {\bibinfo {title} {Radiometerdruck
  und Akkommodationskoeffizient},}\ }\href {\doibase 10.1002/andp.19303980202}
  {\bibfield  {journal} {\bibinfo  {journal} {Ann. Phys.}\ }\textbf {\bibinfo
  {volume} {398}},\ \bibinfo {pages} {129--185} (\bibinfo {year}
  {1930})}\BibitemShut {NoStop}%
\bibitem [{\citenamefont {Martin}(2010)}]{Martin_2010}%
  \BibitemOpen
  \bibfield  {author} {\bibinfo {author} {\bibfnamefont {Holger}\ \bibnamefont
  {Martin}},\ }\bibfield  {title} {\enquote {\bibinfo {title} {Reynolds,
  maxwell, and the radiometer, revisited},}\ }in\ \href@noop {} {\emph
  {\bibinfo {booktitle} {2010 14th International Heat Transfer Conference}}}\
  (\bibinfo {organization} {American Society of Mechanical Engineers},\
  \bibinfo {year} {2010})\ pp.\ \bibinfo {pages} {111--116}\BibitemShut
  {NoStop}%
\bibitem [{\citenamefont {Yang}\ and\ \citenamefont
  {Ripoll}(2014)}]{Yang_2014}%
  \BibitemOpen
  \bibfield  {author} {\bibinfo {author} {\bibfnamefont {Mingcheng}\
  \bibnamefont {Yang}}\ and\ \bibinfo {author} {\bibfnamefont {Marisol}\
  \bibnamefont {Ripoll}},\ }\bibfield  {title} {\enquote {\bibinfo {title} {A
  self-propelled thermophoretic microgear},}\ }\href {\doibase
  10.1039/C3SM52417E} {\bibfield  {journal} {\bibinfo  {journal} {Soft matter}\
  }\textbf {\bibinfo {volume} {10}},\ \bibinfo {pages} {1006--1011} (\bibinfo
  {year} {2014})}\BibitemShut {NoStop}%
\bibitem [{\citenamefont {Wolfe}\ \emph {et~al.}(2016)\citenamefont {Wolfe},
  \citenamefont {Larraza},\ and\ \citenamefont {Garcia}}]{Wolfe_2016}%
  \BibitemOpen
  \bibfield  {author} {\bibinfo {author} {\bibfnamefont {David}\ \bibnamefont
  {Wolfe}}, \bibinfo {author} {\bibfnamefont {Andres}\ \bibnamefont {Larraza}},
  \ and\ \bibinfo {author} {\bibfnamefont {Alejandro}\ \bibnamefont {Garcia}},\
  }\bibfield  {title} {\enquote {\bibinfo {title} {A horizontal vane
  radiometer: Experiment, theory, and simulation},}\ }\href {\doibase
  10.1063/1.4943543} {\bibfield  {journal} {\bibinfo  {journal} {Phys.\
  Fluids}\ }\textbf {\bibinfo {volume} {28}},\ \bibinfo {eid} {037103}
  (\bibinfo {year} {2016}),\ 10.1063/1.4943543}\BibitemShut {NoStop}%
\bibitem [{\citenamefont {Taguchi}\ and\ \citenamefont
  {Aoki}(2015)}]{Taguchi_2015}%
  \BibitemOpen
  \bibfield  {author} {\bibinfo {author} {\bibfnamefont {Satoshi}\ \bibnamefont
  {Taguchi}}\ and\ \bibinfo {author} {\bibfnamefont {Kazuo}\ \bibnamefont
  {Aoki}},\ }\bibfield  {title} {\enquote {\bibinfo {title} {Motion of an array
  of plates in a rarefied gas caused by radiometric force},}\ }\href {\doibase
  10.1103/PhysRevE.91.063007} {\bibfield  {journal} {\bibinfo  {journal}
  {Phys.\ Rev.\ E}\ }\textbf {\bibinfo {volume} {91}},\ \bibinfo {pages}
  {063007} (\bibinfo {year} {2015})}\BibitemShut {NoStop}%
\bibitem [{\citenamefont {Honig}\ and\ \citenamefont
  {Ducker}(2010)}]{Honig_2010}%
  \BibitemOpen
  \bibfield  {author} {\bibinfo {author} {\bibfnamefont {Christopher~DF}\
  \bibnamefont {Honig}}\ and\ \bibinfo {author} {\bibfnamefont {William~A}\
  \bibnamefont {Ducker}},\ }\bibfield  {title} {\enquote {\bibinfo {title}
  {Effect of molecularly-thin films on lubrication forces and accommodation
  coefficients in air},}\ }\href {\doibase 10.1021/jp107106f} {\bibfield
  {journal} {\bibinfo  {journal} {J. Phys. Chem. C}\ }\textbf {\bibinfo
  {volume} {114}},\ \bibinfo {pages} {20114--20119} (\bibinfo {year}
  {2010})}\BibitemShut {NoStop}%
\bibitem [{\citenamefont {Seo}\ and\ \citenamefont {Ducker}(2013)}]{Seo_2013}%
  \BibitemOpen
  \bibfield  {author} {\bibinfo {author} {\bibfnamefont {Dongjin}\ \bibnamefont
  {Seo}}\ and\ \bibinfo {author} {\bibfnamefont {William~A}\ \bibnamefont
  {Ducker}},\ }\bibfield  {title} {\enquote {\bibinfo {title} {In situ control
  of gas flow by modification of gas-solid interactions},}\ }\href {\doibase
  10.1103/PhysRevLett.111.174502} {\bibfield  {journal} {\bibinfo  {journal}
  {Phys.\ Rev.\ Lett.}\ }\textbf {\bibinfo {volume} {111}},\ \bibinfo {pages}
  {174502} (\bibinfo {year} {2013})}\BibitemShut {NoStop}%
\bibitem [{\citenamefont {Seo}\ and\ \citenamefont
  {Ducker}(2014{\natexlab{a}})}]{Seo_2014erratum}%
  \BibitemOpen
  \bibfield  {author} {\bibinfo {author} {\bibfnamefont {Dongjin}\ \bibnamefont
  {Seo}}\ and\ \bibinfo {author} {\bibfnamefont {William~A}\ \bibnamefont
  {Ducker}},\ }\bibfield  {title} {\enquote {\bibinfo {title} {Erratum: In situ
  control of gas flow by modification of gas-solid interactions [Phys.\ Rev.\
  Lett. 111, 174502 (2013)]},}\ }\href {\doibase
  10.1103/PhysRevLett.112.159904} {\bibfield  {journal} {\bibinfo  {journal}
  {Phys.\ Rev.\ Lett.}\ }\textbf {\bibinfo {volume} {112}},\ \bibinfo {pages}
  {159904} (\bibinfo {year} {2014}{\natexlab{a}})}\BibitemShut {NoStop}%
\bibitem [{\citenamefont {Seo}\ and\ \citenamefont
  {Ducker}(2014{\natexlab{b}})}]{Seo_2014}%
  \BibitemOpen
  \bibfield  {author} {\bibinfo {author} {\bibfnamefont {Dongjin}\ \bibnamefont
  {Seo}}\ and\ \bibinfo {author} {\bibfnamefont {William~A}\ \bibnamefont
  {Ducker}},\ }\bibfield  {title} {\enquote {\bibinfo {title} {Effect of gas
  species on gas--monolayer interactions: Tangential momentum accommodation},}\
  }\href {\doibase 10.1021/jp503416x} {\bibfield  {journal} {\bibinfo
  {journal} {J.\ Phys.\ Chem.\ C}\ }\textbf {\bibinfo {volume} {118}},\
  \bibinfo {pages} {20275--20282} (\bibinfo {year}
  {2014}{\natexlab{b}})}\BibitemShut {NoStop}%
\bibitem [{\citenamefont {Miyoshi}\ \emph {et~al.}(2014)\citenamefont
  {Miyoshi}, \citenamefont {Osuka}, \citenamefont {Kinefuchi}, \citenamefont
  {Takagi},\ and\ \citenamefont {Matsumoto}}]{Miyoshi_2014}%
  \BibitemOpen
  \bibfield  {author} {\bibinfo {author} {\bibfnamefont {Nobuya}\ \bibnamefont
  {Miyoshi}}, \bibinfo {author} {\bibfnamefont {Kenichi}\ \bibnamefont
  {Osuka}}, \bibinfo {author} {\bibfnamefont {Ikuya}\ \bibnamefont
  {Kinefuchi}}, \bibinfo {author} {\bibfnamefont {Shu}\ \bibnamefont {Takagi}},
  \ and\ \bibinfo {author} {\bibfnamefont {Yoichiro}\ \bibnamefont
  {Matsumoto}},\ }\bibfield  {title} {\enquote {\bibinfo {title} {Molecular
  beam study of the scattering behavior of water molecules from a graphite
  surface},}\ }\href {\doibase 10.1021/jp500884p} {\bibfield  {journal}
  {\bibinfo  {journal} {J. Phys. Chem. A}\ }\textbf {\bibinfo {volume} {118}},\
  \bibinfo {pages} {4611--4619} (\bibinfo {year} {2014})}\BibitemShut {NoStop}%
\bibitem [{\citenamefont {Lei}\ and\ \citenamefont
  {McKenzie}(2015)}]{Lei_2015}%
  \BibitemOpen
  \bibfield  {author} {\bibinfo {author} {\bibfnamefont {Wenwen}\ \bibnamefont
  {Lei}}\ and\ \bibinfo {author} {\bibfnamefont {David~R}\ \bibnamefont
  {McKenzie}},\ }\bibfield  {title} {\enquote {\bibinfo {title} {Enhanced water
  vapor flow in silica microchannels: The effect of adsorbed water on
  tangential momentum accommodation},}\ }\href {\doibase
  10.1021/acs.jpcc.5b06241} {\bibfield  {journal} {\bibinfo  {journal} {J.
  Phys. Chem. C}\ }\textbf {\bibinfo {volume} {119}},\ \bibinfo {pages}
  {22072--22079} (\bibinfo {year} {2015})}\BibitemShut {NoStop}%
\bibitem [{\citenamefont {Blanchard}\ and\ \citenamefont
  {Ligrani}(2007)}]{Blanchard_2007}%
  \BibitemOpen
  \bibfield  {author} {\bibinfo {author} {\bibfnamefont {Danny}\ \bibnamefont
  {Blanchard}}\ and\ \bibinfo {author} {\bibfnamefont {Phil}\ \bibnamefont
  {Ligrani}},\ }\bibfield  {title} {\enquote {\bibinfo {title} {Slip and
  accommodation coefficients from rarefaction and roughness in rotating
  microscale disk flows},}\ }\href {\doibase 10.1063/1.2739416} {\bibfield
  {journal} {\bibinfo  {journal} {Phys.\ Fluids}\ }\textbf {\bibinfo {volume}
  {19}},\ \bibinfo {pages} {063602} (\bibinfo {year} {2007})}\BibitemShut
  {NoStop}%
\bibitem [{\citenamefont {Cao}\ \emph {et~al.}(2009)\citenamefont {Cao},
  \citenamefont {Sun}, \citenamefont {Chen},\ and\ \citenamefont
  {Guo}}]{Cao_2009}%
  \BibitemOpen
  \bibfield  {author} {\bibinfo {author} {\bibfnamefont {Bing-Yang}\
  \bibnamefont {Cao}}, \bibinfo {author} {\bibfnamefont {Jun}\ \bibnamefont
  {Sun}}, \bibinfo {author} {\bibfnamefont {Min}\ \bibnamefont {Chen}}, \ and\
  \bibinfo {author} {\bibfnamefont {Zeng-Yuan}\ \bibnamefont {Guo}},\
  }\bibfield  {title} {\enquote {\bibinfo {title} {Molecular momentum transport
  at fluid-solid interfaces in MEMS/NEMS: a review},}\ }\href {\doibase
  10.3390/ijms10114638} {\bibfield  {journal} {\bibinfo  {journal} {Int. J.
  Mol. Sci.}\ }\textbf {\bibinfo {volume} {10}},\ \bibinfo {pages} {4638--4706}
  (\bibinfo {year} {2009})}\BibitemShut {NoStop}%
\bibitem [{\citenamefont {Hardt}\ \emph {et~al.}(2009)\citenamefont {Hardt},
  \citenamefont {Tiwari},\ and\ \citenamefont {Klar}}]{Hardt_2009}%
  \BibitemOpen
  \bibfield  {author} {\bibinfo {author} {\bibfnamefont {Steffen}\ \bibnamefont
  {Hardt}}, \bibinfo {author} {\bibfnamefont {Sudarshan}\ \bibnamefont
  {Tiwari}}, \ and\ \bibinfo {author} {\bibfnamefont {Axel}\ \bibnamefont
  {Klar}},\ }\bibfield  {title} {\enquote {\bibinfo {title} {Momentum transfer
  to nanoobjects between isothermal parallel plates},}\ }\href {\doibase
  10.1007/s10404-008-0327-z} {\bibfield  {journal} {\bibinfo  {journal}
  {Microfluid.\ Nanofluid.}\ }\textbf {\bibinfo {volume} {6}},\ \bibinfo
  {pages} {489--498} (\bibinfo {year} {2009})}\BibitemShut {NoStop}%
\bibitem [{\citenamefont {{Bird}}(1994)}]{Bird_1994}%
  \BibitemOpen
  \bibfield  {author} {\bibinfo {author} {\bibfnamefont {G.A.}\ \bibnamefont
  {{Bird}}},\ }\href@noop {} {\emph {\bibinfo {title} {Molecular Gas Dynamics
  and the Direct Simulation of Gas Flows}}}\ (\bibinfo  {publisher} {Clarendon
  Press},\ \bibinfo {year} {1994})\BibitemShut {NoStop}%
\bibitem [{\citenamefont {Scanlon}\ \emph {et~al.}(2010)\citenamefont
  {Scanlon}, \citenamefont {Roohi}, \citenamefont {White}, \citenamefont
  {Darbandi},\ and\ \citenamefont {Reese}}]{Scanlon_2010}%
  \BibitemOpen
  \bibfield  {author} {\bibinfo {author} {\bibfnamefont {TJ}~\bibnamefont
  {Scanlon}}, \bibinfo {author} {\bibfnamefont {E}~\bibnamefont {Roohi}},
  \bibinfo {author} {\bibfnamefont {C}~\bibnamefont {White}}, \bibinfo {author}
  {\bibfnamefont {M}~\bibnamefont {Darbandi}}, \ and\ \bibinfo {author}
  {\bibfnamefont {JM}~\bibnamefont {Reese}},\ }\bibfield  {title} {\enquote
  {\bibinfo {title} {An open source, parallel DSMC code for rarefied gas flows
  in arbitrary geometries},}\ }\href@noop {} {\bibfield  {journal} {\bibinfo
  {journal} {Comput.\ Fluids}\ }\textbf {\bibinfo {volume} {39}},\ \bibinfo
  {pages} {2078--2089} (\bibinfo {year} {2010})}\BibitemShut {NoStop}%
\bibitem [{\citenamefont {Palharini}\ \emph {et~al.}(2015)\citenamefont
  {Palharini}, \citenamefont {White}, \citenamefont {Scanlon}, \citenamefont
  {Brown}, \citenamefont {Borg},\ and\ \citenamefont {Reese}}]{Palharini_2015}%
  \BibitemOpen
  \bibfield  {author} {\bibinfo {author} {\bibfnamefont {Rodrigo~C}\
  \bibnamefont {Palharini}}, \bibinfo {author} {\bibfnamefont {Craig}\
  \bibnamefont {White}}, \bibinfo {author} {\bibfnamefont {Thomas~J}\
  \bibnamefont {Scanlon}}, \bibinfo {author} {\bibfnamefont {Richard~E}\
  \bibnamefont {Brown}}, \bibinfo {author} {\bibfnamefont {Matthew~K}\
  \bibnamefont {Borg}}, \ and\ \bibinfo {author} {\bibfnamefont {Jason~M}\
  \bibnamefont {Reese}},\ }\bibfield  {title} {\enquote {\bibinfo {title}
  {Benchmark numerical simulations of rarefied non-reacting gas flows using an
  open-source DSMC code},}\ }\href@noop {} {\bibfield  {journal} {\bibinfo
  {journal} {Comput. Fluids}\ }\textbf {\bibinfo {volume} {120}},\ \bibinfo
  {pages} {140--157} (\bibinfo {year} {2015})}\BibitemShut {NoStop}%
\bibitem [{\citenamefont {Scanlon}\ \emph {et~al.}(2015)\citenamefont
  {Scanlon}, \citenamefont {White}, \citenamefont {Borg}, \citenamefont
  {Palharini}, \citenamefont {Farbar}, \citenamefont {Boyd}, \citenamefont
  {Reese},\ and\ \citenamefont {Brown}}]{Scanlon_2015}%
  \BibitemOpen
  \bibfield  {author} {\bibinfo {author} {\bibfnamefont {Thomas~J}\
  \bibnamefont {Scanlon}}, \bibinfo {author} {\bibfnamefont {Craig}\
  \bibnamefont {White}}, \bibinfo {author} {\bibfnamefont {Matthew~K}\
  \bibnamefont {Borg}}, \bibinfo {author} {\bibfnamefont {Rodrigo~C}\
  \bibnamefont {Palharini}}, \bibinfo {author} {\bibfnamefont {Erin}\
  \bibnamefont {Farbar}}, \bibinfo {author} {\bibfnamefont {Iain~D}\
  \bibnamefont {Boyd}}, \bibinfo {author} {\bibfnamefont {Jason~M}\
  \bibnamefont {Reese}}, \ and\ \bibinfo {author} {\bibfnamefont {Richard~E}\
  \bibnamefont {Brown}},\ }\bibfield  {title} {\enquote {\bibinfo {title}
  {Open-source direct simulation Monte Carlo chemistry modeling for hypersonic
  flows},}\ }\href {\doibase 10.2514/1.J053370} {\bibfield  {journal} {\bibinfo
   {journal} {AIAA Journal}\ }\textbf {\bibinfo {volume} {53}},\ \bibinfo
  {pages} {1670--1680} (\bibinfo {year} {2015})}\BibitemShut {NoStop}%
\bibitem [{\citenamefont {Weller}\ \emph {et~al.}(1998)\citenamefont {Weller},
  \citenamefont {Tabor}, \citenamefont {Jasak},\ and\ \citenamefont
  {Fureby}}]{Weller_1998}%
  \BibitemOpen
  \bibfield  {author} {\bibinfo {author} {\bibfnamefont {Henry~G}\ \bibnamefont
  {Weller}}, \bibinfo {author} {\bibfnamefont {G}~\bibnamefont {Tabor}},
  \bibinfo {author} {\bibfnamefont {Hrvoje}\ \bibnamefont {Jasak}}, \ and\
  \bibinfo {author} {\bibfnamefont {C}~\bibnamefont {Fureby}},\ }\bibfield
  {title} {\enquote {\bibinfo {title} {A tensorial approach to computational
  continuum mechanics using object-oriented techniques},}\ }\href {\doibase
  10.1063/1.168744} {\bibfield  {journal} {\bibinfo  {journal} {Comput.\
  Phys.}\ }\textbf {\bibinfo {volume} {12}},\ \bibinfo {pages} {620--631}
  (\bibinfo {year} {1998})}\BibitemShut {NoStop}%
\bibitem [{\citenamefont {Ahmad}(2013)}]{Ahmad_2013}%
  \BibitemOpen
  \bibfield  {author} {\bibinfo {author} {\bibfnamefont {Abdul~Ossman}\
  \bibnamefont {Ahmad}},\ }\emph {\bibinfo {title} {Advances in an Open-Source
  Direct Simulation Monte Carlo Technique for Hypersonic Rarefied Gas Flows}},\
  \href {http://ethos.bl.uk/OrderDetails.do?uin=uk.bl.ethos.685090} {Ph.D.
  thesis},\ \bibinfo  {school} {University of Strathclyde} (\bibinfo {year}
  {2013})\BibitemShut {NoStop}%
\bibitem [{\citenamefont {Borgnakke}\ and\ \citenamefont
  {Larsen}(1975)}]{Borgnakke_1975}%
  \BibitemOpen
  \bibfield  {author} {\bibinfo {author} {\bibfnamefont {Claus}\ \bibnamefont
  {Borgnakke}}\ and\ \bibinfo {author} {\bibfnamefont {Poul~S}\ \bibnamefont
  {Larsen}},\ }\bibfield  {title} {\enquote {\bibinfo {title} {Statistical
  collision model for Monte Carlo simulation of polyatomic gas mixture},}\
  }\href {\doibase 10.1016/0021-9991(75)90094-7} {\bibfield  {journal}
  {\bibinfo  {journal} {J. Comput. Phys.}\ }\textbf {\bibinfo {volume} {18}},\
  \bibinfo {pages} {405--420} (\bibinfo {year} {1975})}\BibitemShut {NoStop}%
\bibitem [{\citenamefont {Taguchi}\ and\ \citenamefont
  {Aoki}(2012)}]{Taguchi_2012}%
  \BibitemOpen
  \bibfield  {author} {\bibinfo {author} {\bibfnamefont {Satoshi}\ \bibnamefont
  {Taguchi}}\ and\ \bibinfo {author} {\bibfnamefont {Kazuo}\ \bibnamefont
  {Aoki}},\ }\bibfield  {title} {\enquote {\bibinfo {title} {Rarefied gas flow
  around a sharp edge induced by a temperature field},}\ }\href {\doibase
  10.1017/jfm.2011.536} {\bibfield  {journal} {\bibinfo  {journal} {J.\ Fluid
  Mech.}\ }\textbf {\bibinfo {volume} {694}},\ \bibinfo {pages} {191--224}
  (\bibinfo {year} {2012})}\BibitemShut {NoStop}%
\bibitem [{\citenamefont {Selden}\ \emph {et~al.}(2009)\citenamefont {Selden},
  \citenamefont {Ngalande}, \citenamefont {Gimelshein}, \citenamefont {Muntz},
  \citenamefont {Alexeenko},\ and\ \citenamefont {Ketsdever}}]{Selden_2009a}%
  \BibitemOpen
  \bibfield  {author} {\bibinfo {author} {\bibfnamefont {N}~\bibnamefont
  {Selden}}, \bibinfo {author} {\bibfnamefont {C}~\bibnamefont {Ngalande}},
  \bibinfo {author} {\bibfnamefont {S}~\bibnamefont {Gimelshein}}, \bibinfo
  {author} {\bibfnamefont {EP}~\bibnamefont {Muntz}}, \bibinfo {author}
  {\bibfnamefont {A}~\bibnamefont {Alexeenko}}, \ and\ \bibinfo {author}
  {\bibfnamefont {A}~\bibnamefont {Ketsdever}},\ }\bibfield  {title} {\enquote
  {\bibinfo {title} {Area and edge effects in radiometric forces},}\ }\href
  {\doibase 10.1103/PhysRevE.79.041201} {\bibfield  {journal} {\bibinfo
  {journal} {Phys.\ Rev.\ E}\ }\textbf {\bibinfo {volume} {79}},\ \bibinfo
  {pages} {041201} (\bibinfo {year} {2009})}\BibitemShut {NoStop}%
\bibitem [{\citenamefont {Waldmann}(1959)}]{Waldmann_1959}%
  \BibitemOpen
  \bibfield  {author} {\bibinfo {author} {\bibfnamefont {L}~\bibnamefont
  {Waldmann}},\ }\bibfield  {title} {\enquote {\bibinfo {title} {{\"U}ber die
  Kraft eines inhomogenen Gases auf kleine suspendierte Kugeln},}\ }\href
  {\doibase 10.1515/zna-1959-0701} {\bibfield  {journal} {\bibinfo  {journal}
  {Z.\ Naturforsch.\ Teil A}\ }\textbf {\bibinfo {volume} {14}},\ \bibinfo
  {pages} {589} (\bibinfo {year} {1959})}\BibitemShut {NoStop}%
\bibitem [{\citenamefont {Bakanov}\ and\ \citenamefont
  {Derjaguin}(1960)}]{Bakanov_1960}%
  \BibitemOpen
  \bibfield  {author} {\bibinfo {author} {\bibfnamefont {SP}~\bibnamefont
  {Bakanov}}\ and\ \bibinfo {author} {\bibfnamefont {BV}~\bibnamefont
  {Derjaguin}},\ }\bibfield  {title} {\enquote {\bibinfo {title} {The motion of
  a small particle in a non-uniform gas mixture},}\ }\href {\doibase
  10.1039/DF9603000130} {\bibfield  {journal} {\bibinfo  {journal} {Discuss.
  Faraday Soc.}\ }\textbf {\bibinfo {volume} {30}},\ \bibinfo {pages}
  {130--138} (\bibinfo {year} {1960})}\BibitemShut {NoStop}%
\bibitem [{\citenamefont {Sone}\ and\ \citenamefont
  {Yoshimoto}(1997)}]{Sone_1997}%
  \BibitemOpen
  \bibfield  {author} {\bibinfo {author} {\bibfnamefont {Yoshio}\ \bibnamefont
  {Sone}}\ and\ \bibinfo {author} {\bibfnamefont {Minoru}\ \bibnamefont
  {Yoshimoto}},\ }\bibfield  {title} {\enquote {\bibinfo {title} {Demonstration
  of a rarefied gas flow induced near the edge of a uniformly heated plate},}\
  }\href {\doibase 10.1063/1.869461} {\bibfield  {journal} {\bibinfo  {journal}
  {Phys.\ Fluids}\ }\textbf {\bibinfo {volume} {9}},\ \bibinfo {pages}
  {3530--3534} (\bibinfo {year} {1997})}\BibitemShut {NoStop}%
\bibitem [{\citenamefont {Aoki}\ \emph {et~al.}(1995)\citenamefont {Aoki},
  \citenamefont {Sone},\ and\ \citenamefont {Masukawa}}]{Aoki_1995}%
  \BibitemOpen
  \bibfield  {author} {\bibinfo {author} {\bibfnamefont {Kazuo}\ \bibnamefont
  {Aoki}}, \bibinfo {author} {\bibfnamefont {Yoshio}\ \bibnamefont {Sone}}, \
  and\ \bibinfo {author} {\bibfnamefont {Noboru}\ \bibnamefont {Masukawa}},\
  }\bibfield  {title} {\enquote {\bibinfo {title} {A rarefied gas flow induced
  by a temperature field},}\ }\href@noop {} {\bibfield  {journal} {\bibinfo
  {journal} {Rarefied Gas Dynamics}\ }\textbf {\bibinfo {volume} {1}},\
  \bibinfo {pages} {35--41} (\bibinfo {year} {1995})}\BibitemShut {NoStop}%
\bibitem [{\citenamefont {Lockerby}\ \emph {et~al.}(2004)\citenamefont
  {Lockerby}, \citenamefont {Reese}, \citenamefont {Emerson},\ and\
  \citenamefont {Barber}}]{Lockerby_2004}%
  \BibitemOpen
  \bibfield  {author} {\bibinfo {author} {\bibfnamefont {Duncan~A}\
  \bibnamefont {Lockerby}}, \bibinfo {author} {\bibfnamefont {Jason~M}\
  \bibnamefont {Reese}}, \bibinfo {author} {\bibfnamefont {David~R}\
  \bibnamefont {Emerson}}, \ and\ \bibinfo {author} {\bibfnamefont {Robert~W}\
  \bibnamefont {Barber}},\ }\bibfield  {title} {\enquote {\bibinfo {title}
  {Velocity boundary condition at solid walls in rarefied gas calculations},}\
  }\href {\doibase 10.1103/PhysRevE.70.017303} {\bibfield  {journal} {\bibinfo
  {journal} {Phys. Rev. E}\ }\textbf {\bibinfo {volume} {70}},\ \bibinfo
  {pages} {017303} (\bibinfo {year} {2004})}\BibitemShut {NoStop}%
\bibitem [{\citenamefont {Scandurra}\ \emph {et~al.}(2007)\citenamefont
  {Scandurra}, \citenamefont {Iacopetti},\ and\ \citenamefont
  {Colona}}]{Scandurra_2007}%
  \BibitemOpen
  \bibfield  {author} {\bibinfo {author} {\bibfnamefont {Marco}\ \bibnamefont
  {Scandurra}}, \bibinfo {author} {\bibfnamefont {Fabrizio}\ \bibnamefont
  {Iacopetti}}, \ and\ \bibinfo {author} {\bibfnamefont {Paolo}\ \bibnamefont
  {Colona}},\ }\bibfield  {title} {\enquote {\bibinfo {title} {Gas kinetic
  forces on thin plates in the presence of thermal gradients},}\ }\href
  {\doibase 10.1103/PhysRevE.75.026308} {\bibfield  {journal} {\bibinfo
  {journal} {Phys.\ Rev.\ E}\ }\textbf {\bibinfo {volume} {75}},\ \bibinfo
  {pages} {026308} (\bibinfo {year} {2007})}\BibitemShut {NoStop}%
\bibitem [{\citenamefont {Ventura}\ \emph {et~al.}(2013)\citenamefont
  {Ventura}, \citenamefont {Gimelshein}, \citenamefont {Gimelshein},\ and\
  \citenamefont {Ketsdever}}]{Ventura_2013}%
  \BibitemOpen
  \bibfield  {author} {\bibinfo {author} {\bibfnamefont {Austin}\ \bibnamefont
  {Ventura}}, \bibinfo {author} {\bibfnamefont {Natalia}\ \bibnamefont
  {Gimelshein}}, \bibinfo {author} {\bibfnamefont {Sergey}\ \bibnamefont
  {Gimelshein}}, \ and\ \bibinfo {author} {\bibfnamefont {Andrew}\ \bibnamefont
  {Ketsdever}},\ }\bibfield  {title} {\enquote {\bibinfo {title} {Effect of
  vane thickness on radiometric force},}\ }\href {\doibase
  10.1017/jfm.2013.523} {\bibfield  {journal} {\bibinfo  {journal} {J.\ Fluid
  Mech.}\ }\textbf {\bibinfo {volume} {735}},\ \bibinfo {pages} {684--704}
  (\bibinfo {year} {2013})}\BibitemShut {NoStop}%
\bibitem [{\citenamefont {Agrawal}\ and\ \citenamefont
  {Prabhu}(2008)}]{Agrawal_2008}%
  \BibitemOpen
  \bibfield  {author} {\bibinfo {author} {\bibfnamefont {Amit}\ \bibnamefont
  {Agrawal}}\ and\ \bibinfo {author} {\bibfnamefont {SV}~\bibnamefont
  {Prabhu}},\ }\bibfield  {title} {\enquote {\bibinfo {title} {Survey on
  measurement of tangential momentum accommodation coefficient},}\ }\href
  {\doibase 10.1116/1.2943641} {\bibfield  {journal} {\bibinfo  {journal} {J.
  Vac. Sci. Technol., A}\ }\textbf {\bibinfo {volume} {26}},\ \bibinfo {pages}
  {634--645} (\bibinfo {year} {2008})}\BibitemShut {NoStop}%
\end{thebibliography}
%\input{./Paper_Vane.bbl}
%%%%%%%%%%%%%%%%%%%%%%%%%%%%%%%%%%%%%%%%%%%%%%%%%%%%%%%%
%merlin.mbs apsrev4-1.bst 2010-07-25 4.21a (PWD, AO, DPC) hacked
%Control: key (0)
%Control: author (0) dotless jnrlst
%Control: editor formatted (1) identically to author
%Control: production of article title (0) allowed
%Control: page (1) range
%Control: year (0) verbatim
%Control: production of eprint (0) enabled
%

%%%%%%%%%%%%%%%%%%%%%%%%%%%%%%%%%%%%%%%%%%%%%%%%%%%%%%%%

%\clearpage

\begin{table}
\begin{center}
	\begin {tabular}{cc|c>{\columncolor [gray]{.9}}c|c>{\columncolor [gray]{.9}}c|c>{\columncolor [gray]{.9}}c|c>{\columncolor [gray]{.9}}c}%
Kn&H/W&$-\bar {\tau }_{xy}/p_0$&$\frac {\sigma (\bar {\tau }_{xy})}{|\bar {\tau }_{xy}|}$&$\bar {p}_{yy}/p_0$&$\frac {\sigma (\bar {p}_{yy})}{|\bar {p}_{yy}|}$&$-\dot {m}/\dot {m}_0$&$\frac {\sigma (\dot {m})}{|\dot {m}|}$&$\bar {q}_y/(p_0 c_0)$&$\frac {\sigma (\bar {q}_y)}{|\bar {q}_y|}$\\\rule {0pt}{3ex}%
\pgfutilensuremath {0.005}&\pgfutilensuremath {0.0125}&\pgfutilensuremath {1.95_{-4}}&\pgfutilensuremath {3.3_{-2}}&\pgfutilensuremath {1.72}&\pgfutilensuremath {2.5_{-3}}&\pgfutilensuremath {9.12_{-3}}&\pgfutilensuremath {9.5_{-3}}&\pgfutilensuremath {1.37_{-2}}&\pgfutilensuremath {1.4_{-2}}\\%
\pgfutilensuremath {0.005}&\pgfutilensuremath {0.025}&\pgfutilensuremath {1.28_{-4}}&\pgfutilensuremath {9.5_{-3}}&\pgfutilensuremath {1.66}&\pgfutilensuremath {2.3_{-3}}&\pgfutilensuremath {6.24_{-3}}&\pgfutilensuremath {9.8_{-3}}&\pgfutilensuremath {1.60_{-2}}&\pgfutilensuremath {1.3_{-2}}\\%
\pgfutilensuremath {0.005}&\pgfutilensuremath {0.1}&\pgfutilensuremath {6.19_{-5}}&\pgfutilensuremath {3.6_{-2}}&\pgfutilensuremath {1.55}&\pgfutilensuremath {1.4_{-3}}&\pgfutilensuremath {2.74_{-3}}&\pgfutilensuremath {2.8_{-2}}&\pgfutilensuremath {2.01_{-2}}&\pgfutilensuremath {7.7_{-3}}\\%
\pgfutilensuremath {0.005}&\pgfutilensuremath {0.2}&\pgfutilensuremath {6.05_{-5}}&\pgfutilensuremath {1.9_{-2}}&\pgfutilensuremath {1.48}&\pgfutilensuremath {1.1_{-3}}&\pgfutilensuremath {2.14_{-3}}&\pgfutilensuremath {2.7_{-2}}&\pgfutilensuremath {2.31_{-2}}&\pgfutilensuremath {6.3_{-3}}\\%
\pgfutilensuremath {0.005}&\pgfutilensuremath {0.3}&\pgfutilensuremath {6.99_{-5}}&\pgfutilensuremath {8.5_{-2}}&\pgfutilensuremath {1.41}&\pgfutilensuremath {7.7_{-4}}&\pgfutilensuremath {1.80_{-3}}&\pgfutilensuremath {9.3_{-3}}&\pgfutilensuremath {2.61_{-2}}&\pgfutilensuremath {5.3_{-3}}\\%
\pgfutilensuremath {0.005}&\pgfutilensuremath {0.4}&\pgfutilensuremath {9.01_{-5}}&\pgfutilensuremath {4.2_{-2}}&\pgfutilensuremath {1.36}&\pgfutilensuremath {6.3_{-4}}&\pgfutilensuremath {1.62_{-3}}&\pgfutilensuremath {9.9_{-3}}&\pgfutilensuremath {2.95_{-2}}&\pgfutilensuremath {3.9_{-3}}\\%
\pgfutilensuremath {0.005}&\pgfutilensuremath {0.5}&\pgfutilensuremath {1.21_{-4}}&\pgfutilensuremath {4.0_{-2}}&\pgfutilensuremath {1.31}&\pgfutilensuremath {3.1_{-4}}&\pgfutilensuremath {1.56_{-3}}&\pgfutilensuremath {8.7_{-3}}&\pgfutilensuremath {3.38_{-2}}&\pgfutilensuremath {2.1_{-3}}\\%
\pgfutilensuremath {0.005}&\pgfutilensuremath {0.6}&\pgfutilensuremath {1.74_{-4}}&\pgfutilensuremath {1.3_{-2}}&\pgfutilensuremath {1.26}&\pgfutilensuremath {3.3_{-4}}&\pgfutilensuremath {1.49_{-3}}&\pgfutilensuremath {9.6_{-3}}&\pgfutilensuremath {3.97_{-2}}&\pgfutilensuremath {1.7_{-3}}\\%
\pgfutilensuremath {0.01}&\pgfutilensuremath {0.0125}&\pgfutilensuremath {5.59_{-4}}&\pgfutilensuremath {7.9_{-3}}&\pgfutilensuremath {1.78}&\pgfutilensuremath {1.4_{-3}}&\pgfutilensuremath {1.33_{-2}}&\pgfutilensuremath {6.5_{-3}}&\pgfutilensuremath {2.16_{-2}}&\pgfutilensuremath {7.6_{-3}}\\%
\pgfutilensuremath {0.01}&\pgfutilensuremath {0.025}&\pgfutilensuremath {4.08_{-4}}&\pgfutilensuremath {5.7_{-4}}&\pgfutilensuremath {1.70}&\pgfutilensuremath {1.3_{-3}}&\pgfutilensuremath {1.02_{-2}}&\pgfutilensuremath {6.1_{-3}}&\pgfutilensuremath {2.75_{-2}}&\pgfutilensuremath {7.4_{-3}}\\%
\pgfutilensuremath {0.01}&\pgfutilensuremath {0.1}&\pgfutilensuremath {2.05_{-4}}&\pgfutilensuremath {1.2_{-2}}&\pgfutilensuremath {1.57}&\pgfutilensuremath {9.5_{-4}}&\pgfutilensuremath {4.65_{-3}}&\pgfutilensuremath {1.5_{-2}}&\pgfutilensuremath {3.73_{-2}}&\pgfutilensuremath {4.5_{-3}}\\%
\pgfutilensuremath {0.01}&\pgfutilensuremath {0.2}&\pgfutilensuremath {1.73_{-4}}&\pgfutilensuremath {6.0_{-2}}&\pgfutilensuremath {1.49}&\pgfutilensuremath {1.2_{-3}}&\pgfutilensuremath {3.30_{-3}}&\pgfutilensuremath {5.2_{-3}}&\pgfutilensuremath {4.37_{-2}}&\pgfutilensuremath {5.8_{-3}}\\%
\pgfutilensuremath {0.01}&\pgfutilensuremath {0.3}&\pgfutilensuremath {2.08_{-4}}&\pgfutilensuremath {2.3_{-2}}&\pgfutilensuremath {1.42}&\pgfutilensuremath {5.9_{-4}}&\pgfutilensuremath {2.77_{-3}}&\pgfutilensuremath {1.1_{-2}}&\pgfutilensuremath {4.92_{-2}}&\pgfutilensuremath {3.2_{-3}}\\%
\pgfutilensuremath {0.01}&\pgfutilensuremath {0.4}&\pgfutilensuremath {2.46_{-4}}&\pgfutilensuremath {2.9_{-2}}&\pgfutilensuremath {1.37}&\pgfutilensuremath {2.8_{-4}}&\pgfutilensuremath {2.50_{-3}}&\pgfutilensuremath {6.3_{-3}}&\pgfutilensuremath {5.53_{-2}}&\pgfutilensuremath {1.7_{-3}}\\%
\pgfutilensuremath {0.01}&\pgfutilensuremath {0.5}&\pgfutilensuremath {3.36_{-4}}&\pgfutilensuremath {2.6_{-2}}&\pgfutilensuremath {1.32}&\pgfutilensuremath {1.6_{-4}}&\pgfutilensuremath {2.25_{-3}}&\pgfutilensuremath {4.8_{-3}}&\pgfutilensuremath {6.30_{-2}}&\pgfutilensuremath {1.2_{-3}}\\%
\pgfutilensuremath {0.01}&\pgfutilensuremath {0.6}&\pgfutilensuremath {4.67_{-4}}&\pgfutilensuremath {6.7_{-3}}&\pgfutilensuremath {1.27}&\pgfutilensuremath {2.1_{-4}}&\pgfutilensuremath {2.05_{-3}}&\pgfutilensuremath {6.2_{-3}}&\pgfutilensuremath {7.31_{-2}}&\pgfutilensuremath {1.3_{-3}}\\%
\pgfutilensuremath {0.03}&\pgfutilensuremath {0.5}&\pgfutilensuremath {1.33_{-3}}&\pgfutilensuremath {2.7_{-3}}&\pgfutilensuremath {1.33}&\pgfutilensuremath {3.3_{-5}}&\pgfutilensuremath {3.26_{-3}}&\pgfutilensuremath {7.1_{-4}}&\pgfutilensuremath {1.58_{-1}}&\pgfutilensuremath {1.2_{-4}}\\%
\pgfutilensuremath {0.1}&\pgfutilensuremath {0.0125}&\pgfutilensuremath {5.14_{-3}}&\pgfutilensuremath {2.5_{-3}}&\pgfutilensuremath {1.94}&\pgfutilensuremath {1.5_{-4}}&\pgfutilensuremath {1.43_{-2}}&\pgfutilensuremath {1.3_{-3}}&\pgfutilensuremath {4.29_{-2}}&\pgfutilensuremath {2.0_{-3}}\\%
\pgfutilensuremath {0.1}&\pgfutilensuremath {0.025}&\pgfutilensuremath {6.40_{-3}}&\pgfutilensuremath {1.3_{-3}}&\pgfutilensuremath {1.89}&\pgfutilensuremath {1.3_{-4}}&\pgfutilensuremath {1.75_{-2}}&\pgfutilensuremath {6.3_{-4}}&\pgfutilensuremath {7.61_{-2}}&\pgfutilensuremath {1.4_{-3}}\\%
\pgfutilensuremath {0.1}&\pgfutilensuremath {0.05}&\pgfutilensuremath {7.00_{-3}}&\pgfutilensuremath {9.3_{-4}}&\pgfutilensuremath {1.81}&\pgfutilensuremath {1.1_{-4}}&\pgfutilensuremath {1.83_{-2}}&\pgfutilensuremath {5.0_{-4}}&\pgfutilensuremath {1.24_{-1}}&\pgfutilensuremath {6.6_{-4}}\\%
\pgfutilensuremath {0.1}&\pgfutilensuremath {0.1}&\pgfutilensuremath {6.77_{-3}}&\pgfutilensuremath {1.1_{-3}}&\pgfutilensuremath {1.70}&\pgfutilensuremath {1.1_{-4}}&\pgfutilensuremath {1.60_{-2}}&\pgfutilensuremath {3.5_{-4}}&\pgfutilensuremath {1.83_{-1}}&\pgfutilensuremath {3.1_{-4}}\\%
\pgfutilensuremath {0.1}&\pgfutilensuremath {0.2}&\pgfutilensuremath {5.48_{-3}}&\pgfutilensuremath {7.4_{-4}}&\pgfutilensuremath {1.57}&\pgfutilensuremath {4.7_{-5}}&\pgfutilensuremath {1.05_{-2}}&\pgfutilensuremath {5.7_{-4}}&\pgfutilensuremath {2.46_{-1}}&\pgfutilensuremath {9.5_{-5}}\\%
\pgfutilensuremath {0.1}&\pgfutilensuremath {0.3}&\pgfutilensuremath {4.52_{-3}}&\pgfutilensuremath {1.8_{-3}}&\pgfutilensuremath {1.49}&\pgfutilensuremath {2.5_{-5}}&\pgfutilensuremath {6.77_{-3}}&\pgfutilensuremath {4.3_{-4}}&\pgfutilensuremath {2.85_{-1}}&\pgfutilensuremath {1.1_{-4}}\\%
\pgfutilensuremath {0.1}&\pgfutilensuremath {0.4}&\pgfutilensuremath {4.11_{-3}}&\pgfutilensuremath {2.0_{-3}}&\pgfutilensuremath {1.43}&\pgfutilensuremath {3.6_{-5}}&\pgfutilensuremath {4.64_{-3}}&\pgfutilensuremath {6.0_{-4}}&\pgfutilensuremath {3.17_{-1}}&\pgfutilensuremath {7.2_{-5}}\\%
\pgfutilensuremath {0.1}&\pgfutilensuremath {0.5}&\pgfutilensuremath {4.10_{-3}}&\pgfutilensuremath {3.0_{-3}}&\pgfutilensuremath {1.38}&\pgfutilensuremath {1.5_{-5}}&\pgfutilensuremath {3.31_{-3}}&\pgfutilensuremath {1.2_{-3}}&\pgfutilensuremath {3.46_{-1}}&\pgfutilensuremath {5.9_{-5}}\\%
\pgfutilensuremath {0.1}&\pgfutilensuremath {0.6}&\pgfutilensuremath {4.41_{-3}}&\pgfutilensuremath {2.3_{-3}}&\pgfutilensuremath {1.33}&\pgfutilensuremath {2.3_{-5}}&\pgfutilensuremath {2.40_{-3}}&\pgfutilensuremath {2.2_{-4}}&\pgfutilensuremath {3.77_{-1}}&\pgfutilensuremath {1.9_{-5}}\\%
\pgfutilensuremath {0.3}&\pgfutilensuremath {0.5}&\pgfutilensuremath {8.68_{-3}}&\pgfutilensuremath {1.6_{-3}}&\pgfutilensuremath {1.44}&\pgfutilensuremath {2.6_{-5}}&\pgfutilensuremath {2.28_{-3}}&\pgfutilensuremath {8.5_{-4}}&\pgfutilensuremath {5.23_{-1}}&\pgfutilensuremath {1.6_{-5}}\\%
\pgfutilensuremath {1}&\pgfutilensuremath {0.0125}&\pgfutilensuremath {9.19_{-3}}&\pgfutilensuremath {8.5_{-4}}&\pgfutilensuremath {1.97}&\pgfutilensuremath {6.2_{-5}}&\pgfutilensuremath {5.76_{-3}}&\pgfutilensuremath {3.5_{-3}}&\pgfutilensuremath {4.58_{-2}}&\pgfutilensuremath {2.2_{-3}}\\%
\pgfutilensuremath {1}&\pgfutilensuremath {0.025}&\pgfutilensuremath {1.56_{-2}}&\pgfutilensuremath {3.6_{-4}}&\pgfutilensuremath {1.94}&\pgfutilensuremath {7.3_{-5}}&\pgfutilensuremath {8.53_{-3}}&\pgfutilensuremath {2.6_{-3}}&\pgfutilensuremath {8.70_{-2}}&\pgfutilensuremath {1.0_{-3}}\\%
\pgfutilensuremath {1}&\pgfutilensuremath {0.1}&\pgfutilensuremath {3.26_{-2}}&\pgfutilensuremath {6.0_{-4}}&\pgfutilensuremath {1.80}&\pgfutilensuremath {3.5_{-5}}&\pgfutilensuremath {9.80_{-3}}&\pgfutilensuremath {1.9_{-3}}&\pgfutilensuremath {2.69_{-1}}&\pgfutilensuremath {2.7_{-4}}\\%
\pgfutilensuremath {1}&\pgfutilensuremath {0.2}&\pgfutilensuremath {2.94_{-2}}&\pgfutilensuremath {2.8_{-4}}&\pgfutilensuremath {1.67}&\pgfutilensuremath {4.7_{-5}}&\pgfutilensuremath {5.28_{-3}}&\pgfutilensuremath {1.2_{-3}}&\pgfutilensuremath {4.18_{-1}}&\pgfutilensuremath {9.3_{-5}}\\%
\pgfutilensuremath {1}&\pgfutilensuremath {0.3}&\pgfutilensuremath {2.13_{-2}}&\pgfutilensuremath {8.1_{-4}}&\pgfutilensuremath {1.59}&\pgfutilensuremath {2.6_{-5}}&\pgfutilensuremath {2.48_{-3}}&\pgfutilensuremath {1.1_{-3}}&\pgfutilensuremath {5.14_{-1}}&\pgfutilensuremath {5.9_{-5}}\\%
\pgfutilensuremath {1}&\pgfutilensuremath {0.4}&\pgfutilensuremath {1.53_{-2}}&\pgfutilensuremath {9.1_{-4}}&\pgfutilensuremath {1.53}&\pgfutilensuremath {2.1_{-5}}&\pgfutilensuremath {1.17_{-3}}&\pgfutilensuremath {3.6_{-3}}&\pgfutilensuremath {5.81_{-1}}&\pgfutilensuremath {4.9_{-5}}\\%
\pgfutilensuremath {1}&\pgfutilensuremath {0.5}&\pgfutilensuremath {1.14_{-2}}&\pgfutilensuremath {1.3_{-3}}&\pgfutilensuremath {1.48}&\pgfutilensuremath {1.8_{-5}}&\pgfutilensuremath {5.03_{-4}}&\pgfutilensuremath {2.7_{-2}}&\pgfutilensuremath {6.30_{-1}}&\pgfutilensuremath {2.3_{-5}}\\%
\pgfutilensuremath {1}&\pgfutilensuremath {0.6}&\pgfutilensuremath {8.90_{-3}}&\pgfutilensuremath {2.2_{-3}}&\pgfutilensuremath {1.44}&\pgfutilensuremath {2.1_{-5}}&\pgfutilensuremath {2.23_{-4}}&\pgfutilensuremath {6.9_{-2}}&\pgfutilensuremath {6.69_{-1}}&\pgfutilensuremath {7.4_{-6}}\\%
\pgfutilensuremath {3}&\pgfutilensuremath {0.0125}&\pgfutilensuremath {9.93_{-3}}&\pgfutilensuremath {3.8_{-4}}&\pgfutilensuremath {1.97}&\pgfutilensuremath {7.4_{-5}}&\pgfutilensuremath {3.67_{-3}}&\pgfutilensuremath {3.2_{-3}}&\pgfutilensuremath {4.60_{-2}}&\pgfutilensuremath {2.2_{-3}}\\%
\pgfutilensuremath {3}&\pgfutilensuremath {0.025}&\pgfutilensuremath {1.77_{-2}}&\pgfutilensuremath {6.7_{-4}}&\pgfutilensuremath {1.94}&\pgfutilensuremath {5.8_{-5}}&\pgfutilensuremath {4.94_{-3}}&\pgfutilensuremath {9.0_{-4}}&\pgfutilensuremath {8.79_{-2}}&\pgfutilensuremath {8.3_{-4}}\\%
\pgfutilensuremath {3}&\pgfutilensuremath {0.1}&\pgfutilensuremath {3.83_{-2}}&\pgfutilensuremath {2.9_{-4}}&\pgfutilensuremath {1.80}&\pgfutilensuremath {5.1_{-5}}&\pgfutilensuremath {3.98_{-3}}&\pgfutilensuremath {9.0_{-4}}&\pgfutilensuremath {2.81_{-1}}&\pgfutilensuremath {1.4_{-4}}\\%
\pgfutilensuremath {3}&\pgfutilensuremath {0.2}&\pgfutilensuremath {3.29_{-2}}&\pgfutilensuremath {2.9_{-4}}&\pgfutilensuremath {1.68}&\pgfutilensuremath {2.6_{-5}}&\pgfutilensuremath {1.50_{-3}}&\pgfutilensuremath {7.0_{-3}}&\pgfutilensuremath {4.48_{-1}}&\pgfutilensuremath {4.9_{-5}}\\%
\pgfutilensuremath {3}&\pgfutilensuremath {0.3}&\pgfutilensuremath {2.31_{-2}}&\pgfutilensuremath {6.9_{-4}}&\pgfutilensuremath {1.60}&\pgfutilensuremath {2.2_{-5}}&\pgfutilensuremath {4.69_{-4}}&\pgfutilensuremath {6.5_{-3}}&\pgfutilensuremath {5.54_{-1}}&\pgfutilensuremath {1.3_{-5}}\\%
\pgfutilensuremath {3}&\pgfutilensuremath {0.4}&\pgfutilensuremath {1.61_{-2}}&\pgfutilensuremath {8.1_{-4}}&\pgfutilensuremath {1.54}&\pgfutilensuremath {1.1_{-5}}&\pgfutilensuremath {4.82_{-5}}&\pgfutilensuremath {2.6_{-1}}&\pgfutilensuremath {6.25_{-1}}&\pgfutilensuremath {3.4_{-5}}\\%
\pgfutilensuremath {3}&\pgfutilensuremath {0.5}&\pgfutilensuremath {1.16_{-2}}&\pgfutilensuremath {1.3_{-3}}&\pgfutilensuremath {1.49}&\pgfutilensuremath {2.8_{-5}}&\pgfutilensuremath {-1.07_{-4}}&\pgfutilensuremath {1.0_{-2}}&\pgfutilensuremath {6.74_{-1}}&\pgfutilensuremath {3.1_{-5}}\\%
\pgfutilensuremath {3}&\pgfutilensuremath {0.6}&\pgfutilensuremath {8.62_{-3}}&\pgfutilensuremath {1.5_{-3}}&\pgfutilensuremath {1.45}&\pgfutilensuremath {9.4_{-6}}&\pgfutilensuremath {-1.52_{-4}}&\pgfutilensuremath {2.7_{-2}}&\pgfutilensuremath {7.09_{-1}}&\pgfutilensuremath {2.7_{-5}}\\%
\pgfutilensuremath {10}&\pgfutilensuremath {0.5}&\pgfutilensuremath {1.15_{-2}}&\pgfutilensuremath {1.3_{-3}}&\pgfutilensuremath {1.50}&\pgfutilensuremath {2.2_{-5}}&\pgfutilensuremath {-2.59_{-4}}&\pgfutilensuremath {8.2_{-2}}&\pgfutilensuremath {6.95_{-1}}&\pgfutilensuremath {7.4_{-6}}\\%
\end {tabular}%

	\caption{Numerical results}\label{tab:numRes}
\end{center}
\end{table}

\end{document}